\documentclass[nohyperref]{article}

\usepackage{microtype}
\usepackage{graphicx}
\usepackage{subfigure}
\usepackage{booktabs} 

\usepackage{hyperref}

\usepackage[accepted]{icml2022}

\usepackage{amsmath}
\usepackage{amssymb}
\usepackage{mathtools}
\usepackage{amsthm}

\usepackage[capitalize,noabbrev]{cleveref}

\usepackage{float}

\usepackage{amsmath,amsfonts,bm}

\def\eqref#1{equation~\ref{#1}}

\def\1{\bm{1}}

\DeclareMathAlphabet{\mathsfit}{\encodingdefault}{\sfdefault}{m}{sl}
\SetMathAlphabet{\mathsfit}{bold}{\encodingdefault}{\sfdefault}{bx}{n}

\usepackage[capitalize,noabbrev]{cleveref}

\makeatletter
\renewcommand*{\@fnsymbol}[1]{\ensuremath{\ifcase#1\or \dagger\or *\or \mathsection\or
   \ddagger\or \mathparagraph\or \|\or **\or \dagger\dagger
   \or \ddagger\ddagger \else\@ctrerr\fi}}
\makeatother
   
\definecolor{darkblue}{HTML}{1A254B}
\definecolor{lightblue}{HTML}{A7BED3}
\definecolor{blue}{HTML}{114083}
\definecolor{green}{HTML}{81B5AE}
\definecolor{pink}{HTML}{F2545B}
\definecolor{red}{HTML}{A4243B}

\newcommand{\R}{\mathbb{R}}

\DeclarePairedDelimiter\norm{\lVert}{\rVert}

\renewcommand{\epsilon}{\ensuremath\varepsilon}

\renewcommand{\phi}{\ensuremath{\varphi}}

 \newcommand{\bb}{\mathbf{b}}

 \newcommand{\fb}{\mathbf{f}}
 
 \newcommand{\hb}{\mathbf{h}}

 \newcommand{\mb}{\mathbf{m}}

 \newcommand{\sbb}{\mathbf{s}}

 \newcommand{\vbb}{\mathbf{v}}
 
 \newcommand{\xb}{\mathbf{x}}
 \newcommand{\yb}{\mathbf{y}}
 \newcommand{\zb}{\mathbf{z}}

 \newcommand{\Ab}{\mathbf{A}}
 
 \newcommand{\Cb}{\mathbf{C}}

 \newcommand{\Fb}{\mathbf{F}}
 
 \newcommand{\Hb}{\mathbf{H}}

 \newcommand{\Ob}{\mathbf{O}}

 \newcommand{\Sbb}{\mathbf{S}}
 
 \newcommand{\Ub}{\mathbf{U}}
 
 \newcommand{\Wb}{\mathbf{W}}
 \newcommand{\Xb}{\mathbf{X}}
 \newcommand{\Yb}{\mathbf{Y}}
 \newcommand{\Zb}{\mathbf{Z}}

\newcommand{\Ecal}{\mathcal{E}}

\newcommand{\Ncal}{\mathcal{N}}

\newcommand{\Vcal}{\mathcal{V}}

\usepackage{arydshln}

\usepackage{xcolor}
\definecolor{cite_color}{HTML}{114083}
\definecolor{link_color}{RGB}{153, 0,0}  
\definecolor{url_color}{RGB}{153, 102,  0}
\definecolor{emp_color}{RGB}{0,0,255}
\hypersetup{
 colorlinks,
 citecolor=cite_color,
 linkcolor=link_color,
 urlcolor=url_color}
 
\usepackage{etoc}
\etocdepthtag.toc{mtchapter}
\etocsettagdepth{mtchapter}{subsection}
\etocsettagdepth{mtappendix}{none}

\icmltitlerunning{\textsc{EquiBind}: Geometric Deep Learning for Drug Binding Structure Prediction}

\begin{document}

\twocolumn[
\icmltitle{\textsc{EquiBind}: Geometric Deep Learning for Drug Binding Structure Prediction}



\icmlsetsymbol{equal}{*}

\begin{icmlauthorlist}
\icmlauthor{Hannes Stärk}{equal,yyy}
\icmlauthor{Octavian-Eugen Ganea}{equal,yyy}
\icmlauthor{Lagnajit Pattanaik}{yyy}
\icmlauthor{Regina Barzilay}{yyy}
\icmlauthor{Tommi Jaakkola}{yyy}
\end{icmlauthorlist}
\icmlaffiliation{yyy}{Massachusetts Institute of Technology, MIT, Cambridge, MA, USA}

\icmlcorrespondingauthor{Hannes Stärk}{hstark@mit.edu}

\icmlkeywords{Machine Learning, ICML}

\vskip 0.3in
]



\printAffiliationsAndNotice{\icmlEqualContribution} 

\begin{abstract}
Predicting how a drug-like molecule binds to a specific protein target is a core problem in drug discovery. An extremely fast computational binding method would enable key applications such as fast virtual screening or drug engineering. Existing methods are computationally expensive as they rely on heavy candidate sampling coupled with scoring, ranking, and fine-tuning steps. We challenge this paradigm with \textsc{EquiBind}, an SE(3)-equivariant geometric deep learning model performing direct-shot prediction of both i) the receptor binding location (blind docking) and ii) the ligand's bound pose and orientation. EquiBind achieves significant speed-ups and better quality compared to traditional and recent baselines. Further, we show extra improvements when coupling it with existing fine-tuning techniques at the cost of increased running time. Finally, we propose a novel and fast fine-tuning model that adjusts torsion angles of a ligand's rotatable bonds based on closed-form global minima of the von Mises angular distance to a given input atomic point cloud, avoiding previous expensive differential evolution strategies for energy minimization.
\end{abstract}

\begin{figure*}
\vspace{-0cm}
\begin{center}
\centerline{\includegraphics[width=1.9\columnwidth]{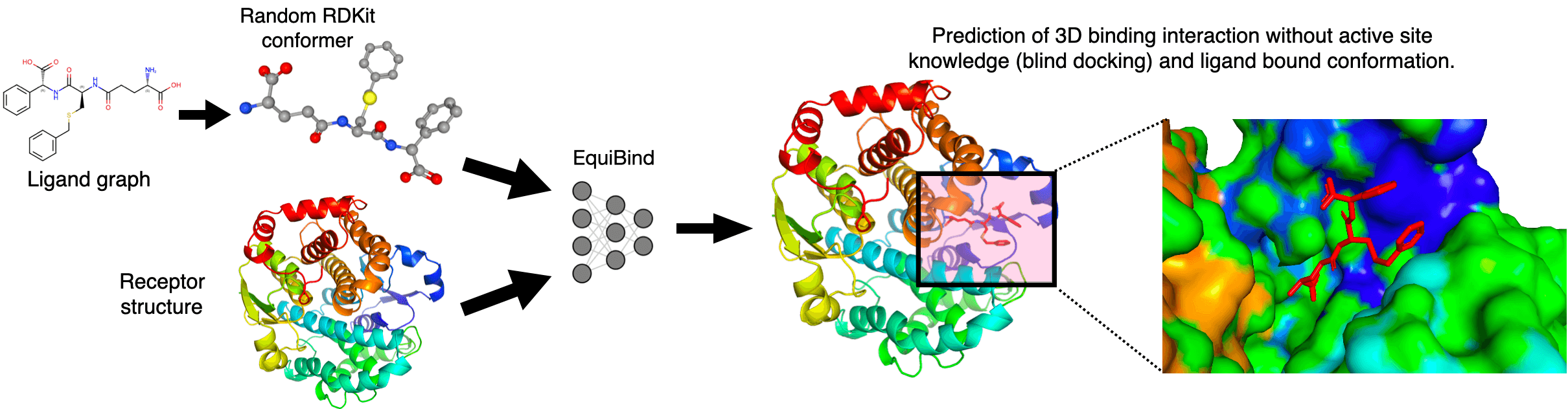}}
\vspace{-0.3cm}
\caption{High-level overview of the structural drug binding problem tackled by \textsc{EquiBind}.}
\label{fig:intro}
\end{center}
 \vskip -0.37in
\end{figure*}

\section{Introduction} \label{sec:intro}

Drug discovery is an expensive process, e.g., a single drug costs around 1 billion dollars and takes 10 years of development and testing before potentially being FDA approved. Moreover, this process can fail at any point, e.g., due to unforeseen side effects or experimental disproof of the promised therapeutic efficacy. Worse, there are  $~10^{60}$ possible drug-like molecules~\citep{reymond2012exploring}, going far beyond current experimental capabilities. 

\par Accurate computational methods, e.g., deep learning (DL) based, can drastically reduce the molecular search space, but need to be extremely fast to scan the vast biological and chemical spaces for both desired and unexpected effects. For instance, a novel drug that inactivates an important cancer protein might negatively inhibit other essential proteins in the human body, potentially resulting in life-threatening  side effects. Given that the human proteome contains up to $100\,000$ protein types, the current hope is to scan for these interactions in a computational manner before bringing a few promising candidates to \textit{in vitro} and \textit{in vivo} testing.

\par A core problem in drug discovery is understanding how drug-like molecules (ligands) interact and form complexes with target proteins (receptors) -- \textit{drug binding} -- which is a prerequisite for virtual screening. This is a difficult problem with different facets and constraints: binding kinetics, conformational changes (internal molecular flexibility), and chemical and geometrical atomic interaction types are part of the domain knowledge describing ligand-protein binding mechanisms~\citep{du2016insights}. For instance, classical models for molecular complex formation are  "lock-and-key", "induced fit", and "conformational selection", while hydrophobic,  hydrogen-bonding, and $\pi$-stacking are the most frequent atomic binding interactions, but other types often occur during binding~\citep{de2017systematic}.


\par Current \textit{in silico} approaches for  (3D) structural drug binding achieve high quality at a significant computational cost: in our experiments the GNINA method~\citep{mcnutt2021gnina} takes on average 146 seconds for a single ligand-receptor pair, while the popular commercial software Glide~\citep{halgren2004glide} is up to 9 times slower. This is caused by the common strategy employed by all previous binding methods: first, a large set of candidate complexes (e.g., millions) is generated via thorough sampling of possible binding locations and poses~\citep{hassan2017protein}; then, scoring and ranking steps are used to retrieve the most promising instances; finally, an energy-based fine-tuning method is employed to best fit the ligand in the respective pocket locations. 

\par Here, we introduce \textsc{EquiBind}, a novel geometric \& graph deep learning model for structural drug binding -- \cref{fig:intro}. Inspired by \citet{ganea2021independent}, we exploit graph matching networks (GMN)~\citep{li2019graph} and E(3)-equivariant graph neural networks (E(3)-GNN)~\citep{satorras2021n} to perform a direct prediction of the ligand-receptor complex structure without relying on heavy sampling as prior work, thus achieving \textbf{significant inference time speed-ups}. Moreover, since 3D structural data suffers from scarcity (e.g., only around 19K experimental complexes are publicly available in the PDBbind database), it is crucial to inject the right physical, chemical, or biological \textit{inductive biases} into DL models to avoid learning these priors from insufficient amounts of data and to create trustable models. Towards this goal, \textsc{EquiBind}:
\vspace{-0.3cm}
\begin{itemize}
    \itemsep .1pt
    \item guarantees independence to the initial 3D placements and orientations of the two molecules, i.e., the exact same complex is always predicted for the same input unbound structures,
    \item incorporates an efficient mechanism for biologically plausible ligand flexibility by only altering torsion angles of rotatable bonds while keeping local structures (bond angles and lengths) fixed,
    \item utilizes a non-intersection loss to prevent steric clashes or unrealistic van der Waals interactions.
\end{itemize}
\vspace{-0.2cm}

We focus on the \textit{blind docking} scenario, i.e., zero knowledge of the protein's binding site or pocket. However, our method can easily be adapted to situations where the approximate binding location is known. Similar to~\citep{zhang2020edock}, we argue that errors in the ground truth binding pocket conformation heavily affect traditional docking methods that are conditioned on the receptor active site~\citep{lang2009dock,trott2010autodock}. In practice, the ground truth 3D locations of the binding atoms might be low-resolution, might not be known at all (e.g., for novel antigens), or we might be interested in discovering new druggable locations on a protein's surface that were previously thought to be undruggable (e.g. exploring allosteric binding locations rather than orthosteric sites).

Empirically, we investigate two  settings: re-docking (i.e., taking the bound ligand structure out of a complex and asking the model to dock it) and flexible self-docking (i.e., ligands have no bound structure knowledge prior to docking). We assume a rigid receptor, but we model ligand flexibility by first predicting an atomic point cloud of the deformed molecule and then employing a fast algorithm to extract internal changes of rotatable bonds' torsion angles that would match the point cloud as well as possible. Instead of minimizing the root-mean-square deviation (RMSD) using expensive optimization strategies (e.g.,  differential
evolution approaches~\citep{mendez2021geometric}), we maximize the log-likelihood of a von Mises distribution that fits the torsion angles, proving closed-form expressions of the global optimum. Experimentally, we show improved quality in various metrics over popular and recent state-of-the-art baselines at a fraction of the running time. Finally, we show the power of combining EquiBind with existing energy-based methods to realize a hybrid DL approach. Indeed, we believe the future of computational drug discovery will follow the paradigm demonstrated here.

\section{Related Work} \label{sec:rltd_work}

\textbf{Protein and molecular structure prediction. } Obtaining experimental 3D structural data of molecules and proteins is a highly expensive process. However, very recent DL models have produced a breakthrough in computational protein folding~\citep{jumper2021highly,baek2021accurate} and fast generation of small molecule low-energy conformation ensembles~\citep{ganea2021geomol,luo2021predicting,xu2020learning,shi2021learning, yuanqi}. These methods aim to accelerate discovering structures and complement experimental data in various applications such as drug discovery.

\textbf{Protein representations (for DL-based molecular interactions). } To be useful for predicting molecular interactions, proteins must be modeled in specific ways to account for different views: backbone \& side-chains, protein \mbox{surface}, atomic point cloud, or amino-acid sequence. \citet{somnath2021multi} create a hierarchical representation of proteins and prove its utility in binding and function prediction. \citet{gainza2020deciphering,sverrisson2021fast} leverage geometric deep learning and mesh convolutional neural networks (CNN) to embed protein surface patches into fingerprints and allow for fast scanning and binding site identification, removing the need for handcrafted or expensive pre-computed features. However, these methods do not perform the full structural blind docking task that involves prediction of the binding site, of the orientations and placements of the two molecular structures, and of the internal conformational deformations during binding. Various other protein representations have been proposed for (graph) DL methods for individual structure prediction~\citep{jing2020learning}, protein-protein interactions~\citep{dai2021protein,eismann2020hierarchical,townshend2019end}, or protein function prediction~\citep{gligorijevic2021structure}.

\textbf{Druggable binding site identification.} Traditional computational methods for scanning proteins for their most "druggable" areas have leveraged various views such as utilizing the protein's 3D structure or/and residue sequence, extracting geometric features, building large template libraries, or relying on energy-based models~\citep{macari2019computational}. Recently, DL changed this paradigm, e.g., using 3D CNNs~\citep{aggarwal10deeppocket,jimenez2017deepsite,torng2019high} or sequence models~ \citep{sankararaman2010active}. 

\textbf{Popular and more recent drug binding models.} Representative docking software for drug-like molecules are AutoDock Vina~\citep{trott2010autodock} and its various extensions for improving speed~\citep{trott2010autodock}, scoring~\citep{koes2013lessons} or for blind docking~\citep{hassan2017protein}. As mentioned in \cref{sec:intro}, these methods employ a multi-stage strategy based on heavy candidate sampling, scoring, ranking, and fine-tuning. Various subsequent methods improved parts of this pipeline~\citep{zhang2020edock,mohammad2021instadock,mcnutt2021gnina,francoeur2020three}, such as scoring functions (see below) or by incorporating prior knowledge of the ligand and receptor when bound to other molecules~\citep{Lam2018, Lam2019}.

\textbf{GNNs and CNNs for binding scoring functions and binding affinity prediction.} Deep learning on 3D voxel images (via 3D CNNs) or interaction graphs (via GNNs) have improved the traditional hand-designed scoring function used in AutoDock Vina, enabling better fine-tuning of predicted docked poses, as well as direct binding affinity prediction from the 3D complex \citep{mcnutt2021gnina,francoeur2020three,ragoza2017protein,wallach2015atomnet,lim2019predicting,morrone2020combining,jiang2021interactiongraphnet,shen2021impact,jski2020emulating,bao2021deepbsp,torng2019graph,li2021structure}. However, some methods~\citep{karimi2019deepaffinity,gao2018interpretable} have found that using the protein sequence and the drug SMILES string already provide competitive predictions for binding affinity without the need for 3D structural data.  

Closer to our approach,  \citet{mendez2021geometric} has shown that optimizing the ligand's global 3D position and orientation and the torsion angles of rotatable bonds to minimize a GNN based scoring function improves fine-tuning of the ligand into the active site and its predicted bound pose. However, they employ differential evolution for this optimization, and we find it to be slow in practice (e.g., 29 sec for a 5-rotatable bond molecule and 50 min for a 44-rotatable bond molecule). Instead, our proposed \textsc{EquiBind} optimizes in closed form a ligand's torsion angles to match a predicted atomic point cloud in less than 1 sec. 

\textbf{Applications of drug binding methods.} Computational docking methods are employed for various facets of drug discovery, e.g., fast virtual screening~\citep{gniewek2021learning,jski2020emulating} or de novo binder generation~\citep{masuda2020generating,imrie2021deep,drotar2021structure}.

\textbf{Deep learning for protein-protein docking.} A related problem is \textit{protein-protein docking} in which recent methods have performed direct prediction of the complex structure from the two concatenated input sequences using evolutionary information~\citep{evans2021protein}, or have leveraged geometric deep learning to model rigid body docking~\citep{ganea2021independent} or side-chains structures~\citep{jindal2021side}. 

\textbf{Incorporating Euclidean symmetries into GNNs.} Injecting Euclidean 3D transformations into geometric DL models has become possible using equivariant message passing layers \citep{cohen2016group,thomas2018tensor,fuchs2020se,satorras2021n,brandstetter2021geometric,batzner2021se}. Our method follows \citet{ganea2021independent} to incorporate SE(3) pairwise equivariance into message passing neural networks for the drug binding problem. However, different from this method, we go beyond rigid docking and model ligand conformational flexibility.

\section{\textsc{EquiBind} Model} \label{sec:method}
\begin{figure*}
\vspace{-0.1cm}
\begin{center}
\centerline{\includegraphics[width=1.9\columnwidth]{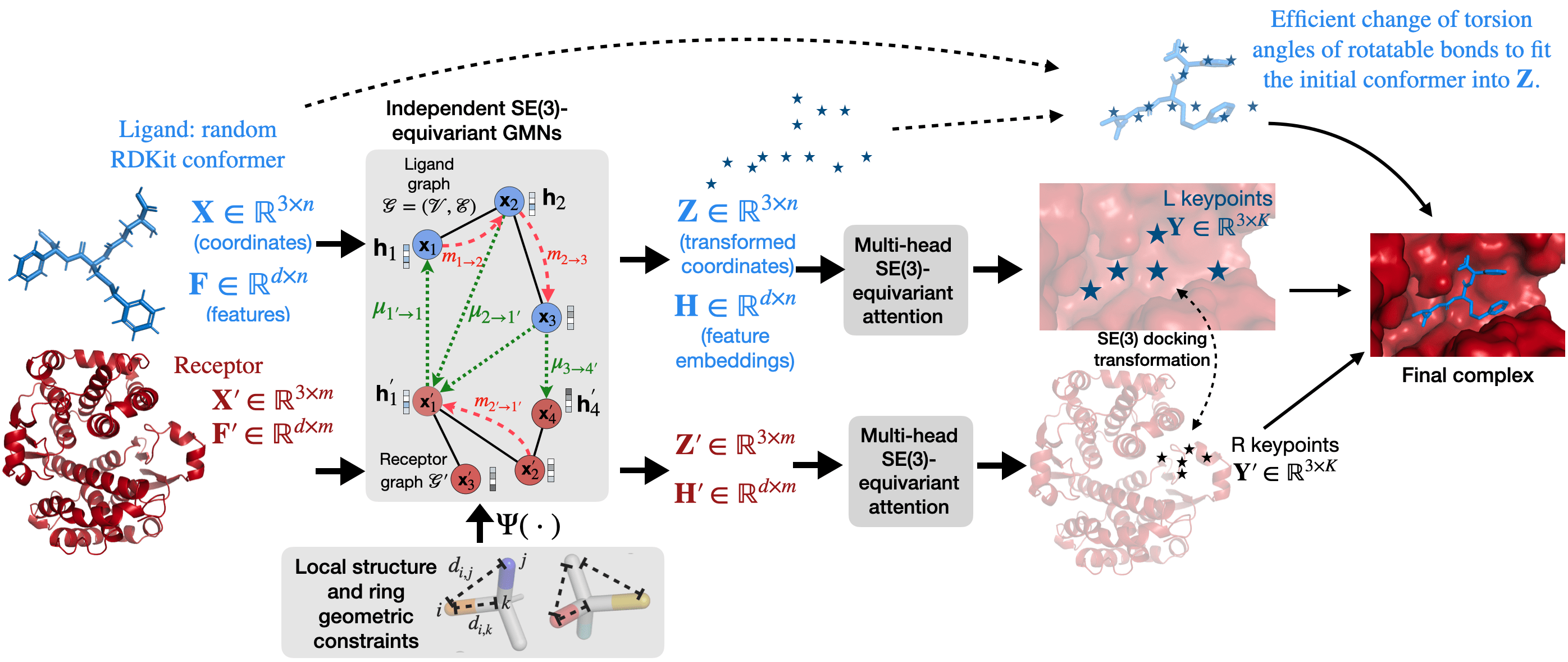}}
\vspace{-0.3cm}
\caption{\textsc{EquiBind} model architecture.}
\label{fig:model}
\end{center}
 \vskip -0.35in
\end{figure*}

We describe our \textsc{EquiBind} model, highlighted in \cref{fig:intro} and detailed in \cref{fig:model}. It takes as input a ligand molecular graph with a random associated unbound 3D conformer (e.g., generated using RDKit/ETKDG \cite{riniker2015better}), as well as a receptor-bound structure. As previously noted, we only model ligand flexibility in this work, assuming a rigid protein  conformation.

\textbf{K-NN graph representations. } We represent both input molecules as spatial k-nearest neighbor (k-NN) graphs. The ligand graph $\mathcal{G} = (\mathcal{V}, \mathcal{E})$ uses atoms as nodes with their respective 3D coordinates from the unbound conformer denoted as $\Xb \in \R^{3 \times n}$, and initial features $\Fb \in \R^{d \times n}$ (e.g., atom type). Edges include all atom pairs within a  distance cutoff of 4 \AA. The receptor graph $\mathcal{G}' = (\mathcal{V}', \mathcal{E}')$ has residues as nodes and their 3D coordinates $\Xb' \in \R^{3 \times m}$ are given by the $\alpha$-carbon locations. Each node is connected in the graph to the closest 10 other nodes at less than 30 \AA\  distance. The receptor node features $\Fb' \in \R^{d \times m}$ and the ligand features are detailed in \cref{appendix:implementation}.

\textbf{Independent E(3)-equivariant transformations.} Similar to~\citep{ganea2021independent}, an important geometric inductive bias is to predict the same binding complex no matter how the initial molecules are positioned and oriented in space. This is especially needed for data-scarce problems such as structural drug binding. Towards this goal, we use \emph{Independent E(3)-Equivariant Graph Matching Network} (IEGMN)~\citep{ganea2021independent} which combines Graph Matching Networks~\citep{li2019graph} and E(3)-Equivariant Graph Neural Networks~\citep{satorras2021n}. This architecture jointly transforms both features and 3D coordinates to perform \textit{intra} and \textit{inter} neural graph message passing. Formally, $\text{IEGMN}(\Xb,\Fb,\Xb',\Fb') = \Zb \in\R^{3\times n},\Hb \in\R^{d\times n},\Zb' \in\R^{3\times m},\Hb' \in\R^{d\times m}$, where $\Zb,\Zb'$ are transformed coordinates, while $\Hb,\Hb'$ are feature embeddings. The core property of IEGMNs is that stacking any number of such layers guarantees that any independent rotation and translation of the original input structures will be exactly reflected in the outputs, i.e., $\text{IEGMN}(\Ub \Xb + \bb,\Fb,\Ub' \Xb' + \bb',\Fb') = \Ub \Zb + \bb,\Hb,\Ub' \Zb' + \bb',\Hb'$ for any orthogonal matrices $\Ub,\Ub' \in \Sbb\Ob(3)$ and translation vectors $\bb,\bb' \in \R^3$. In practice, the $\Zb,\Hb,\Zb',\Hb'$ outputs shown in \cref{fig:model} are obtained by stacking several IEGMN layers. Our choice for a single \textit{l}-th layer is:
$$\mb_{j\to i} = \phi^e(\hb_i^{(l)},\hb_j^{(l)},\norm{\xb_i^{(l)}-\xb_j^{(l)}}^2,\fb_{j\to i}),\forall (i,j)\in\Ecal\cup\Ecal'$$
$$\mu_{j'\to i} = a_{j'\to i} \Wb \hb_{j'}^{(l)},\forall i\in\Vcal,j'\in\Vcal' \text{ or } i\in \Vcal', j'\in\Vcal$$
$$\mb_i = \frac{1}{|\Ncal(i)|} \sum_{j\in\Ncal(i)}\mb_{j\to i},\forall i\in\Vcal\cup\Vcal'$$
$$\mu_{i} =  \sum_{j'\in\Vcal'}\mu_{j'\to i},\forall i \in\Vcal, \quad \text{and} \quad \mu'_{i} =  \sum_{j\in\Vcal} \mu_{j\to i'},\forall i \in\Vcal'$$
$$\xb_i^{(l+1)} = \Psi\left(\xb_i^{(l)} + \sum_{j\in\Ncal(i)}\frac{\xb_i^{(l)}-\xb_j^{(l)}}{\|\xb_i^{(l)}-\xb_j^{(l)} \|}\phi^x(\mb_{j\to i})\right)$$
$$\hb_i^{(l+1)} = (1 - \beta) \cdot \hb_i^{(l)} + \beta \cdot \phi^h(\hb_i^{(l)},\mb_i,\mu_{i}, \fb_i),\forall i\in\Vcal\cup\Vcal'$$
where $a_{j\to i}$ are SE(3)-invariant attention coefficients derived from $\Hb$ embeddings, $\Ncal(i)$ are the graph neighbors of node $i$, $\Wb$ is a parameter matrix, and the various $\phi$ functions are modeled using shallow neural networks, with $\phi^x$ outputting a scalar and others a d-dimensional vector.

When modeling ligand flexibility, we found it useful to incorporate additional geometric constraints on transformed coordinates through $\Psi$ models described in \cref{sssec:dg}.

\textbf{The role of $\Zb$.} The output of the coordinate E(3)-equivariant transformations denoted as $\Zb$ and $\Zb'$ will be used in different roles: to identify the rigid body transformation and the binding site, as well as to model ligand flexibility by training $\Zb$ to represent the deformed atomic point cloud. We detail both steps below.

\subsection{Rigid transformation through binding keypoints} 
To identify the rigid SE(3) transformation to dock the ligand in the right position and orientation, we follow~\citep{ganea2021independent} and compute ligand and receptor keypoints of size $K$ using an SE(3)-equivariant multi-head attention mechanism defined as $\yb_{k} := \sum_{i=1}^n \alpha_i^k \zb_{i}, \quad \yb'_{k} := \sum_{j=1}^m \beta_j^k \zb'_{j}$
where $\alpha_i^k = softmax_{i}(\frac{1}{\sqrt{d}}\hb_{1i}^{\top} \Ub  \mu(\phi(\Hb_2)))$ and similarly defined $\beta_j^k$ are attention coefficients, with $\Ub$ a parametric learnable matrix. These keypoints are trained to match the ground truth binding pocket points using an optimal transport loss that recovers their alignment (detailed in~\citep{ganea2021independent}). In our case, ground truth binding pocket points are defined as midpoints of segments connecting ligand atoms to receptor atoms (e.g., from side-chains) that are closer than 4 \AA\ . For models incorporating ligand flexibility, these pocket points are defined as all ligand atoms that are closer than 4 \AA\ to any receptor atom. When the ligand and receptor are separated, we seek to identify the corresponding binding sites, and their exact matching using the two predicted keypoint sets $\Yb, \Yb' \in \R^{3xK}$. If predicted perfectly, the SE(3) transformation to superimpose $\Yb$ and  $\Yb'$ would precisely correspond to the binding SE(3) transformation to perform ligand docking.

\subsection{Modeling Chemically Plausible Ligand Flexibility}

It has previously been assumed that the most flexible parts of drug-like molecules are rotatable bonds, while \textit{local atomic structures (LAS)} (bond lengths and adjacent bond angles) or small rings are mostly rigid \citep{trott2010autodock,zsoldos2007ehits,huang2018comprehensive,mendez2021geometric}. We here follow this hypothesis in two different ways as below.

We model ligand flexibility through $\Zb$, which will represent a good approximation of the deformed atomic point cloud of the original conformer (i.e., into its bound structure). We train our model with two extra loss function terms: ligand-RMSD (Root-mean-square deviation) and Kabsch-RMSD~\footnote{RMSD after superimposition, or RMSD of atomic positions.}.

\subsubsection{Distance Geometric Constraints}\label{sssec:dg}

Our first goal is to enforce LAS distance constraints in the IEGMN layers after each coordinate transformation, i.e., through a tailored differentiable function $\Psi$, which we call \textit{"LAS distance geometry (DG) projection."} While a hard constraint might be difficult to impose exactly, we find the following soft strategy to work well. Formally, the transformed coordinates $\Zb$ satisfy the LAS DG constraints if they are a global minimum of the following function for a fixed (random low-energy unbound) conformer $\Xb$:
\begin{equation*}\label{equation:dg}
\begin{split}
\mathcal{S}(\Zb, \Xb) & = \ \sum_{\{(i,j) \in \mathcal{E}\}} (d^2_{\Xb}(i,j) - d^2_{\Zb}(i,j))^2 \\
& + \sum_{\{i,j: \text{2-hops\ away\ in\ } \mathcal{G}\}} (d^2_{\Xb}(i,j) - d^2_{\Zb}(i,j))^2 \\
& + \sum_{\{i,j: \text{i in aromatic ring with j\}}} (d^2_{\Xb}(i,j) - d^2_{\Zb}(i,j))^2
\end{split}
\end{equation*}
where $d_{\Xb}(i,j) = \|x_i - x_j \|$. Thus, our definition of $\Psi$ is hard-coding a fixed number (T) of gradient descent layers that aim to minimize $\mathcal{S}$:
\begin{equation*}
    \Psi(\Zb) = \Psi_T \circ \ldots \circ \Psi_1(\Zb), \quad \Psi_t(\Zb) = \Zb - \eta \nabla_{\Zb} \mathcal{S}(\Zb, \Xb), \forall t
\end{equation*}
which is easy since gradients of $\mathcal{S}$ have a simple closed-form expression. A similar approach can be employed for modeling various other rigid substructures such as aromatic rings. T and the correction step size $\eta$ are model hyperparameters chosen as described in Appendix \ref{appendix:implementation}.

\subsubsection{Fast Point Cloud Ligand Fitting}\label{sssec:conformer-fitting}
However, while helpful for model training, the previous gradient descent-based projection is not guaranteed to enforce hard LAS DG constraints and, thus, might produce implausible conformers in practice as we show in \cref{fig:conformer-fitting}.

\par To address this issue, we only change the torsion angles of the initial (RDKit) unbound conformer $\Xb$ to match $\Zb$ as well as possible while keeping LAS fixed and, thus, hard-guaranteeing chemically plausible output bound conformers. The output will be a new conformer $\Cb \in \R^{3xn}$ with $\mathcal{S}(\Cb, \Xb) = 0$. First, $\Cb$ is initialized as $\Xb$, and only its rotatable bonds' torsion angles are changed.

\par A choice is to optimize $\Cb$ for minimizing \mbox{Kabsch-RMSD$(\Zb,\Cb)$}. However, such an approach requires an iterative optimization strategy of all torsion angles of rotatable bonds, which can be done using a differential evolution algorithm as in~\citet{mendez2021geometric}, or other local random search strategies. This is computationally expensive (e.g., 51 minutes for a single 44 rotatable bond molecule) and might fail to find a good local minimum. A gradient-based method could better capture the various molecular interactions, but computing the gradients of a point cloud w.r.t. its bonds' torsion angles is non-trivial given the geometric dependencies between dihedral angles, i.e., \cref{eq:torsion_constraint}.

\par We present a much cheaper alternative for which a closed form solution exists (does not require optimization): we compute the dihedral angles of rotatable bonds of $\Cb$ as maximum likelihood estimates of von Mises distributions on dihedral angles of $\Zb$. Formally, this reduces to the following maximization:
\begin{equation}
    \max_{\{\angle(kij, ijl)\}}  \sum_{(k,i),(i,j),(j,l) \in \mathcal{E}} \cos( \angle_{\Zb}(kij, ijl) - \angle(kij, ijl) )
    \label{eq:dihedral_loss}
\end{equation}
where $\angle_{\Zb}(kij, ijl)$ are the dihedral angles\footnote{We use clockwise angles following the chemistry convention.} of $\Zb$, and $\angle(kij, ijl)$ are the dihedrals of $\Cb$ that we seek to optimize. 

However, we have to explicitly take into account that \textit{all the dihedral angles for the same rotatable bond (i,j) are coupled} by the following constraint~\citep{ganea2021geomol}:
\begin{equation}
    \begin{split}
    & \angle(kij, ijl) =_{2\pi} \angle(k'ij, ijl') + \angle(kij, k'ij) + \angle(ijl', ijl),\\
    & \quad \forall (i,j) \in \mathcal{E}, \forall k,k' \in \mathcal{N}_i, \forall l,l' \in \mathcal{N}_j 
    \label{eq:torsion_constraint}
    \end{split}
\end{equation}
where $\angle(kij, k'ij)$ and  $\angle(ijl', ijl)$ depend only on the local structures of nodes $i$ and $j$, respectively, thus will not change if the torsion angle of bond (i,j) changes. 

To minimize \cref{eq:dihedral_loss}, we can simply do it independently per each rotatable bond $(i,j) \in \mathcal{E}$. Let us fix one such bond $(i,j)$ and use the notations: $\Delta_{kl} = \angle(kij, ijl)$ and $\beta_{kk'l'l} = \angle(kij, k'ij) + \angle(ijl', ijl)$. Additionally, for any angle $\alpha$, we define:
 $\Ab_{\alpha} := \begin{bmatrix}
    \cos(\alpha) & -\sin(\alpha) \\
    \sin(\alpha) & \cos(\alpha)  
\end{bmatrix}$
and  $\sbb_{\alpha} := \begin{bmatrix}
    \cos(\alpha) \\
    \sin(\alpha)   
\end{bmatrix}$. Thus, we rewrite the constraint in \cref{eq:torsion_constraint} as $\sbb_{\Delta_{kl}} = \Ab_{\beta_{kk'l'l}} \sbb_{\Delta_{k'l'}}, \forall k,k' \in \mathcal{N}_i, \forall l,l' \in \mathcal{N}_j$. \cref{eq:dihedral_loss} is then rewritten for bond (i,j), up to a constant:

\begin{equation}
    \max_{\{\Delta_{kl}\}}  \sum_{k \in \mathcal{N}_i}\sum_{l \in \mathcal{N}_j} \langle \sbb_{\Delta_{kl}}, \sbb^*_{\Delta_{kl}} \rangle
\end{equation}
Choosing any fixed $k_0 \in \mathcal{N}_i$, $l_0 \in \mathcal{N}_j$, the above becomes
\begin{equation}
    \max_{\Delta_{k_0l_0}}  \sum_{k \in \mathcal{N}_i}\sum_{l \in \mathcal{N}_j} \langle \Ab_{\beta_{kk_0l_0l}} \sbb_{\Delta_{k_0l_0}}, \sbb^*_{\Delta_{kl}} \rangle = \sbb_{\Delta_{k_0l_0}}^{\top} \vbb  
\end{equation}
where $\vbb := \sum_{k \in \mathcal{N}_i}\sum_{l \in \mathcal{N}_j}  \Ab_{\beta_{kk_0l_0l}}^{\top} \sbb^*_{\Delta_{kl}} $. This has the closed form solution $\sbb_{\Delta_{k_0l_0}} = \frac{\vbb}{\|\vbb\|}$, which finally gives all dihedral angles $\Delta_{kl}$ in closed form. One can easily verify that the choice of $k_0 \in \mathcal{N}_i$ and $l_0 \in \mathcal{N}_j$ will not affect the values of the predicted dihedrals $\Delta_{kl}$, $\forall k \in \mathcal{N}_i, \forall l \in \mathcal{N}_j$. 

In a practical example, the above solution recovers all 44 rotatable bond torsion angles of a randomly modified conformer in  0.04 seconds as opposed to 3143 seconds needed by a differential evolution method.


\setlength{\tabcolsep}{1.8pt}
\begin{table*}[htb]
    \caption{\textbf{Flexible blind self-docking.} All methods receive a random RDKit conformer of the ligand molecule as input and are tasked to find its binding site and the right orientation + conformation in which it binds. \textsc{EquiBind}-U refers to uncorrected ligands $\Zb$ that are not necessarily chemically plausible. \textsc{EquiBind} corrects them with our fast conformer fitting corrections -- see \cref{sssec:conformer-fitting}. \textsc{EquiBind+Q} predicts an approximate ligand position and fine-tunes it using QuickVina 2. \textsc{EquiBind+Q2} samples more candidate positions, and \textsc{EquiBind+S} uses SMINA for fine-tuning. Standard deviations are in Appendix Table \ref{tab:stdevs}. *GLIDE runtime details in Appendix \ref{appendix:implementation}. }
    \label{tab:self-docking}
     \begin{small}
     \begin{center}
     \begin{sc}
     \makebox[\textwidth][c]{
    \begin{tabular}{lcc|cccccc|cccccc|cc}
    
    \toprule
     & & &\multicolumn{6}{c}{Ligand RMSD $\downarrow$} & \multicolumn{6}{c}{Centroid Distance $\downarrow$} & \multicolumn{2}{c}{Kabsch}\\
     &\begin{tabular}{@{}c@{}}\ avg.\\sec.\end{tabular} & \begin{tabular}{@{}c@{}}\ avg.\\sec.\end{tabular} &\multicolumn{4}{c}{Percentiles $\downarrow$} & \multicolumn{2}{c}{\begin{tabular}{@{}c@{}}\% below\\threshold $\uparrow$\end{tabular} }  &\multicolumn{4}{c}{Percentiles $\downarrow$} & \multicolumn{2}{c}{\begin{tabular}{@{}c@{}}\% below\\thresh. $\uparrow$\end{tabular} } & \multicolumn{2}{c}{RMSD $\downarrow$}\\
    
    \textbf{Methods} & 16-CPU & GPU & Mean & 25th & 50th & 75th & 5 \AA{}  &  2 \AA{} & Mean & 25th & 50th & 75th & 5 \AA{}  &  2 \AA{} & Mean & Med.\\
    \midrule
     \textsc{QVina-W}& 49& -& 13.6 & 2.5  &  7.7 &  23.7 & 40.2 & 20.9 & 11.9  &  0.9 &  3.7 & 22.9 & 54.6 & 41.0 & \textbf{2.1} & 1.9\\
     \textsc{GNINA}& 247& 146& 13.3 & 2.8 & 8.7   & 22.1 & 37.1  & 21.2 & 11.5 & 1.0 &  4.5 & 21.2 & 52.0 & 36.0 & 2.2 & 1.8\\
    \textsc{SMINA}&146& - & 12.1 & 3.8 & 8.1 & 17.9 & 33.9 & 13.5 & 9.8 & 1.3 &  3.7 & 16.2 & 55.9 & 38.0 & 2.2 & 1.9\\
    \textsc{GLIDE} (c.)& 1405*& - & 16.2 & 2.6 & 9.3   & 28.1 & 33.6 & 21.8 & 14.4& \textbf{0.8} &  5.6 & 26.9 & 48.7 & 36.1 & 2.2 & 1.9\\
    \hline
    \textsc{EquiBind} &\textbf{0.16}& \textbf{0.04} & \textbf{8.2} & 3.8 & 6.2 &  \textbf{10.3} & 39.1 & 5.5 & \textbf{5.6} & 1.3 & 2.6 & 7.4 & 67.5&  40.0 & 2.6 & 2.3\\
     \textsc{EquiBind+Q}& 8& 8 & 8.4 & 2.6 & 6.6 &  11.1 & 38.0 & 18.7 &  5.9 & 1.0 & 2.5 &  6.4 & 68.7 & 44.6  &2.3 & 1.9\\
     \textsc{EquiBind+Q2}& 15& 15 & 8.7 & 2.6 & 6.8 &  11.1 & 40.7 & 21.6 &  6.0 & 1.0  & 2.4 &  6.6 & 70.1 & 42.7  & 2.2 & \textbf{1.6}\\
     \textsc{EquiBind+S}& 146& 146 & 8.3 & \textbf{2.1} & \textbf{5.6} &  10.5 & \textbf{46.4} & \textbf{24.6} &  6.0 & 0.9  & \textbf{2.0} &  \textbf{6.2} & \textbf{71.0} & \textbf{50.6}  & \textbf{2.1}& 1.8\\
     
     \midrule
     \textsc{EquiBind}-U &0.14& 0.02 & 7.8 & 3.3 & 5.7 &  9.7 & 42.4 & 7.2 & 5.6 & 1.3 & 2.6 & 7.4 & 67.5&  40.0 & 2.1 & 1.8\\
     \bottomrule
    \end{tabular}}
    \end{sc}
    \end{center}
    \end{small}
     \vskip -0.25in
\end{table*}

\section{Experiments}  \label{sec:exps}
\subsection{Data}
We provide a new time-based dataset split and preprocessing pipeline for DL drug binding methods\footnote{We make this data and associated scripts available at \url{https://github.com/HannesStark/EquiBind}.}. We use protein-ligand complexes from PDBBind \cite{liu2017PDBBind}, which is a subset of the Protein Data Bank (PDB) \cite{berman2002PDB} that provides 3D structures of individual proteins and complexes. The newest version, PDBBind v2020, contains $19\,443$ protein-ligand complexes with $3890$ unique receptors and $15\,193$ unique ligands. Histograms for individual receptor and ligand data frequencies are in Figure \ref{dataset-histograms} and we describe our preprocessing to remove pathologies of the data in \cref{appendix:dataset}.

\textbf{Motivation for new test set and time split.} Docking methods are often evaluated using the PDBBind core set, which contains 285 hand-curated high-resolution complexes. However, this might not reflect the performance in real-world applications where  data might not be of similar high quality. Due to the differences in resolution and the average ligand size (32 heavy atoms in PDBBind versus 24 in the core set), the complexes of the core set can be considered easier to predict than the average complex. Moreover, some of the previous methods might have been validated or trained  on a subset of the core set and thus, report optimistic quality numbers. To better reflect the average complex encountered in applications, we employ a test set that only contains complexes that were discovered in 2019 or later, while the train and validation sets only use strictly older complexes. 

\textbf{Dataset split.} Of the $19\,119$ preprocessed complexes, $1512$ were discovered in 2019 or later. From these, we randomly sample $125$ unique proteins and collect all new complexes containing them ($363$) to create the final test set. The low number of test samples is chosen to make it feasible to compare with time-consuming classical physics-based docking methods. From the remaining complexes that are older than $2019$, we remove those with ligands contained in the test set, giving $17\,347$ complexes for training and validation. These are divided into $968$ validation complexes, which share no ligands with the remaining $16\,379$ train complexes. Results when only testing on new receptors are in Appendix \ref{appendix:results}.

\subsection{Evaluation Setup}\label{eval_setup}

\textbf{Baselines.} Quick Vina-W (QVina-W) is a classical docking program specifically developed for "wide" or blind docking. 
SMINA~\citep{koes2013lessons} builds on AutoDock Vina by designing an improved and empirical scoring function. GNINA~\citep{mcnutt2021gnina,francoeur2020three} further develops a DL scoring function using CNNs and a grid-based featurization scheme. GLIDE~\citep{halgren2004glide} is a popular commercial docking software of which we use the 2021-4 release.
We run GLIDE, GNINA, and SMINA with their default settings and for QVina-W we increase the exhaustiveness (parameter controlling the search time) to 64 with which it is still faster than the other baselines.

\textbf{EquiBind models.} Our model can be applied in various scenarios, see caption of \cref{tab:self-docking}. First, the \textsc{EquiBind}-U model generates an uncorrected ligand point cloud $\Zb$ that does not necessarily have valid bond angles and lengths. The standard \textsc{EquiBind} takes this output and applies our fast point cloud ligand fitting in \cref{sssec:conformer-fitting} to obtain a realistic molecular structure. The model \textsc{EquiBind}-R treats the ligand as a rigid body, being trained with no flexibility loss terms. The fine-tuning model \textsc{EquiBind + Q} builds on top of this output by searching refined conformations using Quick Vina 2 in a 5 \AA{}\ bounding box around the ligand predicted by \textsc{EquiBind}-R. The instantiations \textsc{EquiBind + Q2} does the same with two times as many sampled ligand positions, and \textsc{EquiBind + S} as well as \textsc{EquiBind-R + S} instead use SMINA for fine-tuning.

\textbf{Evaluation Metrics.} We use the ligand root mean square deviation (L-RMSD), the centroid distance, and the Kabsch-RMSD. We calculate all metrics after hydrogens are removed. The \textit{centroid distance} measures the ability of the model to find the correct binding pocket (for a given ligand) via the distance between the averaged coordinates of the predicted and true bound ligand atoms. \textit{Kabsch RMSD} is the lowest possible RMSD that can be obtained by SE(3) transformation of the ligand (i.e., RMSD after superimposition with the Kabsch algorithm). \textit{L-RMSD} is the mean squared error between the atoms of the predicted and bound ligands. All RMSDs are calculated using OpenBabel's symmetry corrected and atom order invariant RMSD tool \texttt{obrms}. 
Next to the mean and cumulative distributions, we report the percent of predictions below a given error threshold.

Finally, we show the average number of seconds needed to make a prediction for a test complex, given the receptor and the initial ligand structure. The receptor preparation time is excluded (mainly an additional 393 sec for GLIDE) since this step is commonly only performed once before docking many ligands to the same receptor. We ran all runtime measurements on the same machine using 16 logical CPU cores (except for GLIDE, which does not support multithreading -- detailed in Appendix \ref{appendix:implementation}), once with and once without access to a 6GB GTX 1060 GPU.

\textbf{Implementation Details.} We optimize our model using Adam~\citep{kingma2014adam} and do early stopping with patience of 150 epochs based on the percentage of predicted validation set complexes with an RMSD  better than 2  \AA{}. All hyperparameters and the employed ligand and node features are described in Appendix \ref{appendix:implementation}. Code to reproduce results or perform fast docking with the provided model weights is available at \url{https://github.com/HannesStark/EquiBind}.

\subsection{Results}

\textbf{Blind self-docking.}
This set of experiments reflects the performance that can be expected in the most typical applications where the true ligand bond angles and distances (which are used in re-docking) are unknown. An initial approximate conformer has to be obtained from a 2D molecular graph for which we use a random RDKit conformer. 
\begin{figure}[!h]
\vspace{-0.cm}
\begin{center}
\centerline{\includegraphics[width=.9\columnwidth]{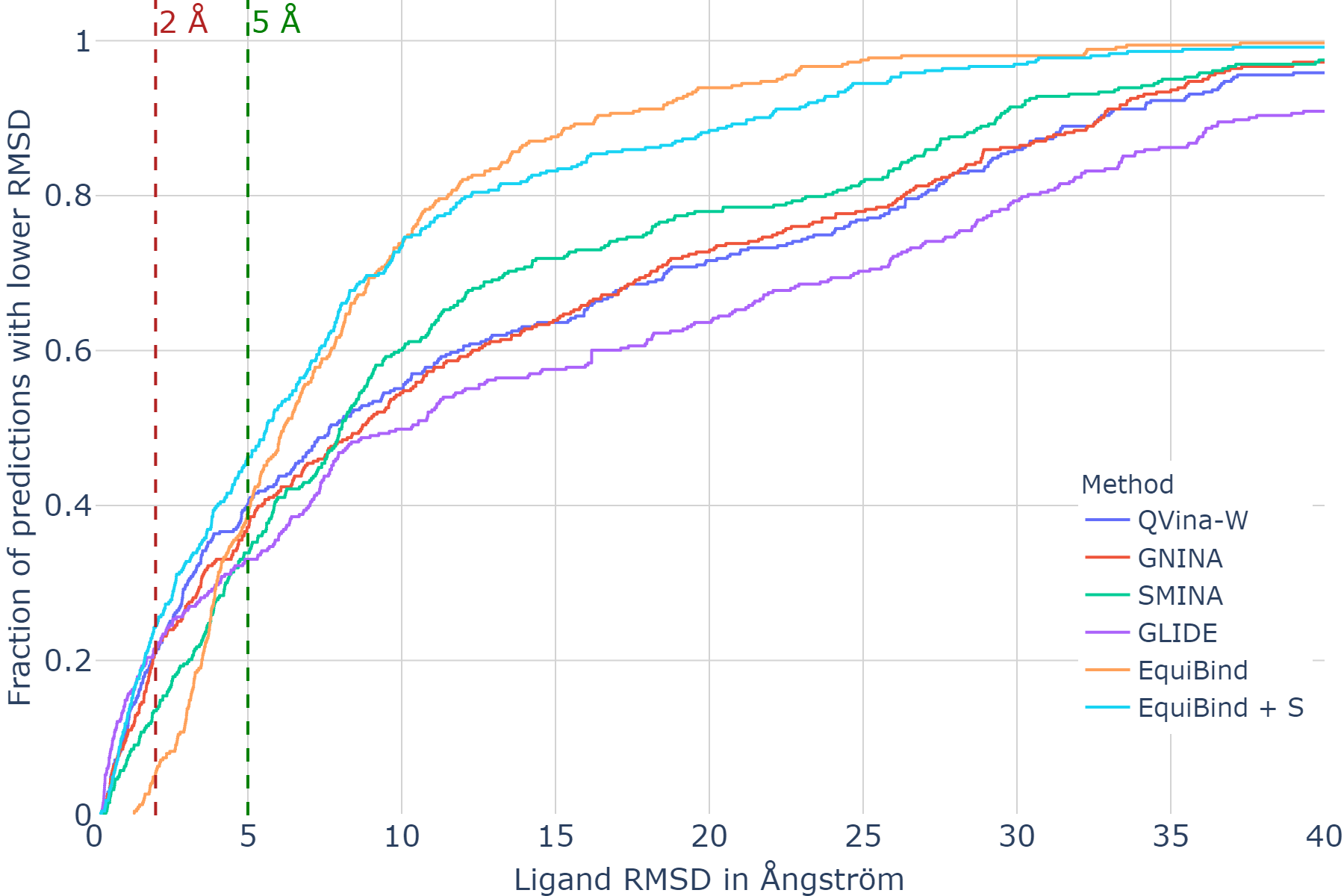}}
\centerline{\includegraphics[width=.9\columnwidth]{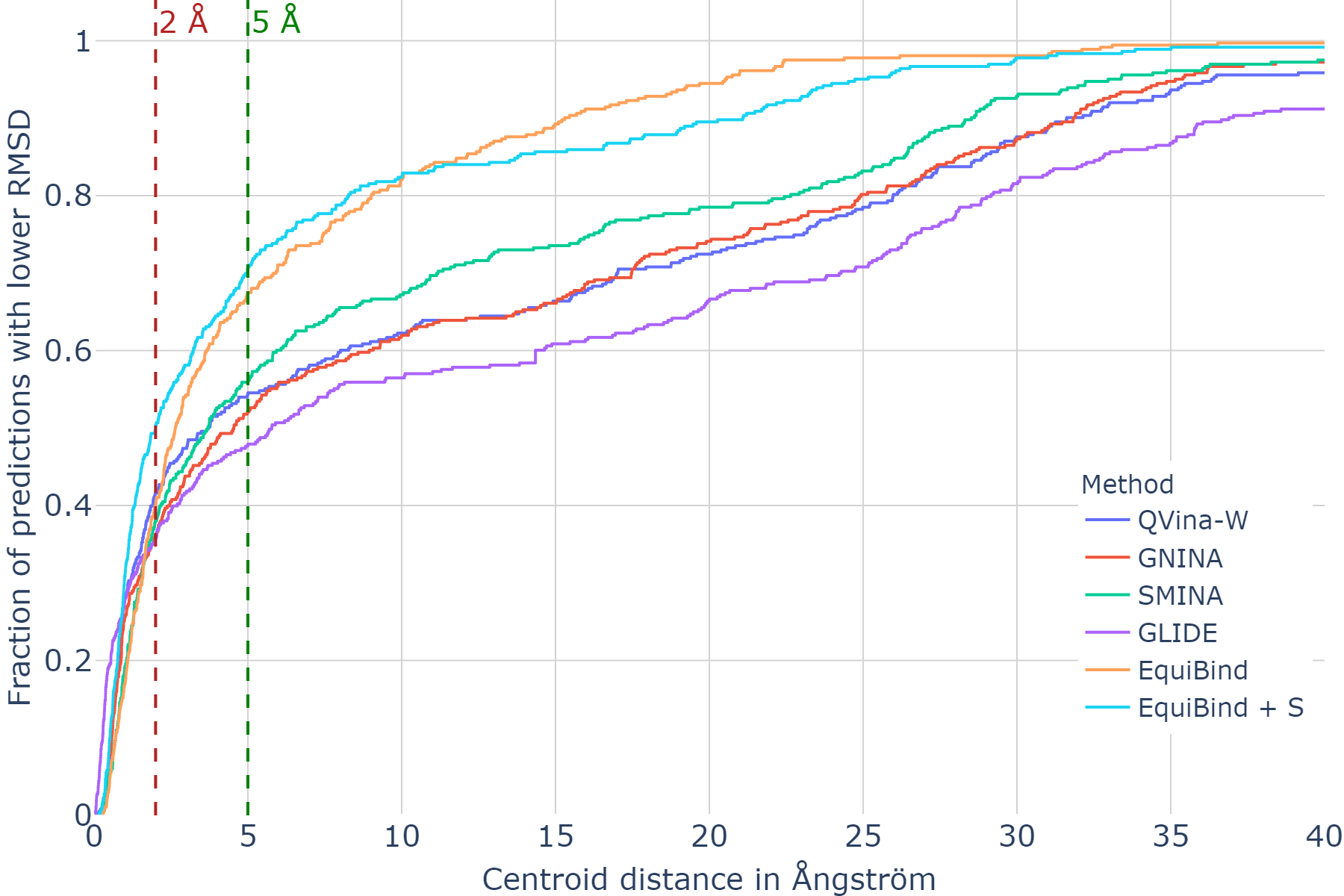}}
\vspace{-0.3cm}
\caption{\textbf{Flexible blind self-docking.} Cumulative density histogram of the L-RMSD (top) and centroid distance (bottom) of \textsc{EquiBind} with and without SMINA for fine-tuning.} 
\label{fig:histograms}
\end{center}
 \vskip -0.4in
\end{figure}

\setlength{\tabcolsep}{1.8pt}
\begin{table*}[htb]
    \caption{\textbf{Blind re-docking.} The input is the bound ligand structure at a random location. The methods are tasked to find the binding site and the right binding location + orientation. \textsc{EquiBind}-R's and GLIDE's Kabsch RMSD is 0 since they treat the ligand as a rigid body while the other methods can only be run in a flexible mode where torsion angles are changed. *GLIDE runtime details in Appendix \ref{appendix:implementation}.}
    \label{tab:re-docking}

     \begin{small}
     \begin{center}
     \begin{sc}
     \makebox[\textwidth][c]{
    \begin{tabular}{lcc|cccccc|cccccc|cc}
    \toprule
     & & &\multicolumn{6}{c}{Ligand RMSD $\downarrow$} & \multicolumn{6}{c}{Centroid Distance $\downarrow$} & \multicolumn{2}{c}{Kabsch}\\
     &\begin{tabular}{@{}c@{}}\ avg.\\sec.\end{tabular} & \begin{tabular}{@{}c@{}}\ avg.\\sec.\end{tabular} &\multicolumn{4}{c}{Percentiles $\downarrow$} & \multicolumn{2}{c}{\begin{tabular}{@{}c@{}}\% below\\threshold $\uparrow$\end{tabular} }  &\multicolumn{4}{c}{Percentiles $\downarrow$} & \multicolumn{2}{c}{\begin{tabular}{@{}c@{}}\% below\\thresh. $\uparrow$\end{tabular} } & \multicolumn{2}{c}{RMSD $\downarrow$}\\
    \textbf{Methods} & 16-CPU & GPU & Mean & 25th & 50th & 75th & 5 \AA{}  &  2 \AA{} & Mean & 25th & 50th & 75th & 5 \AA{}  &  2 \AA{} & Mean & Med.\\
    \midrule
     \textsc{QVina-W}& 49& -& 13.4& 1.6  &  7.9 &  24.1 & 39.0 & 27.7 & 11.8  &  0.9 &  3.8 & 23.2 &  55.4 &  40.4 & 1.8 & 1.5\\
     \textsc{GNINA}& 247& 146& 12.2 & 1.3 & 6.1   & 22.9 & 46.8  & 32.2 & 10.9 & 0.7 &  2.8 & 22.1 & 58.4 & 43.8 & 1.7& 1.4\\
     \textsc{SMINA}&146& - & 10.3 & 1.4 & 6.2 & 15.2 & 46.7 & 30.1 & 8.5 & 0.8 &  2.6 & 12.7 & 63.5 & 45.3 & 1.7 & 1.4\\
     \textsc{GLIDE}& 1405*& - & 15.7 & \textbf{0.5} & 8.3 & 29.5 & 45.7 & \textbf{43.4} & 14.8 & \textbf{0.3} &  4.9 & 28.5 & 50.4 & 45.4 & 0 & 0\\
      \midrule
     \textsc{EquiBind}-R &\textbf{0.14}& \textbf{0.02}& {7.4} & 2.0 & {5.1} &  {9.8} & {49.0} & 25.1 & {5.8} & 1.4 & {2.6} & {7.3} & {66.9}&  40.8 & 0 & 0\\
     \textsc{EquiBind-R+S}& 146& 146 & \textbf{7.0} & 1.0 & \textbf{3.4}   & \textbf{9.6} & \textbf{57.0}  & 41.1 & \textbf{5.3} & 0.7 &  \textbf{1.4} & \textbf{4.7} & \textbf{76.0} & \textbf{59.2} & 1.5& 1.1\\
     \bottomrule
    \end{tabular}}
    \end{sc}
    \end{center}
    \end{small}
    \vskip -0.25in
\end{table*}

The results in Table \ref{tab:self-docking} show that vanilla \textsc{EquiBind} performs well at identifying the approximate binding location and outperforms the baselines in metrics other than the 25th RMSD percentile and the fraction of predictions with and RMSD better than 2 \AA{}. The fine-tuning extensions of \textsc{EquiBind} such as \textsc{EquiBind + Q}  outperform or match the baselines in all metrics, while \textsc{EquiBind + Q} and \textsc{EquiBind + Q2} also retain significant inference speed-ups, making our method \textit{suitable for extremely high-throughput applications such as virtual screening over databases of hundred million molecules, e.g., ZINC}. Thus, practitioners can combine our method with previous fine-tuning baselines and trade quality over runtime depending on the downstream task of interest.

Figure \ref{fig:histograms} shows the same trend for the \textbf{RMSD}. \textsc{EquiBind}, which is three orders of magnitude faster than the fastest baseline, improves over the baselines for the predictions in the $>4$ \AA{}  regime. \textsc{EquiBind} does better for complexes that are hard to predict (e.g., due to ligand size) and also outperforms the baselines in the low RMSD regime when using fine-tuning (\textsc{EquiBind+S}).
For the \textbf{centroid distances}, the exact conformer is less crucial, and the methods mainly have to find the correct binding pocket location. Here, \textsc{EquiBind} is already able to match the baselines in the low error regime without fine-tuning. Histograms for \textsc{EquiBind + Q} and \textsc{EquiBind + Q2} are in Appendix Figure \ref{fig:additional-histogram}. 

The \textbf{main observations} are that \textsc{EquiBind} is much faster than the baselines, has fewer predictions that are far off from the true conformer, and can use fast fine-tuning for very low-RMSD final predictions. The benefits through fine-tuning can be expected considering the difficulty of predicting the correct torsions jointly with the binding location and orientation in a single forward pass.

\textbf{Blind re-docking.}
In these experiments, the bound ligand is extracted from the binding pocket, placed in a random location, and the methods have to re-dock it into the correct conformation. Thus the methods have access to the ground truth structure of the ligand, and all predictions will have the correct bond lengths and angles. \textsc{EquiBind}-R treats the ligand as a completely rigid body and only predicts a translation and rotation. 
Rigid re-docking results are of practical relevance for docking strategies where large amounts of conformations are generated for a single molecule and then rigidly docked to the receptor before using an additional scoring function to rank the predictions.

In \cref{tab:re-docking} we can observe that \textsc{EquiBind}-R can be particularly impactful for this strategy due to its much faster inference time. This is while outperforming the baselines in the metrics other than the 25th percentiles and the fraction of predictions with an error below 2 \AA{}. For practical rigid re-docking applications, this could potentially be remedied by docking 10 times as many conformers while still retaining a 10 times speed-up over the fastest baseline.

\begin{figure}[htb]
\vspace{0.cm}
\begin{center}
\centerline{\includegraphics[width=.9\columnwidth]{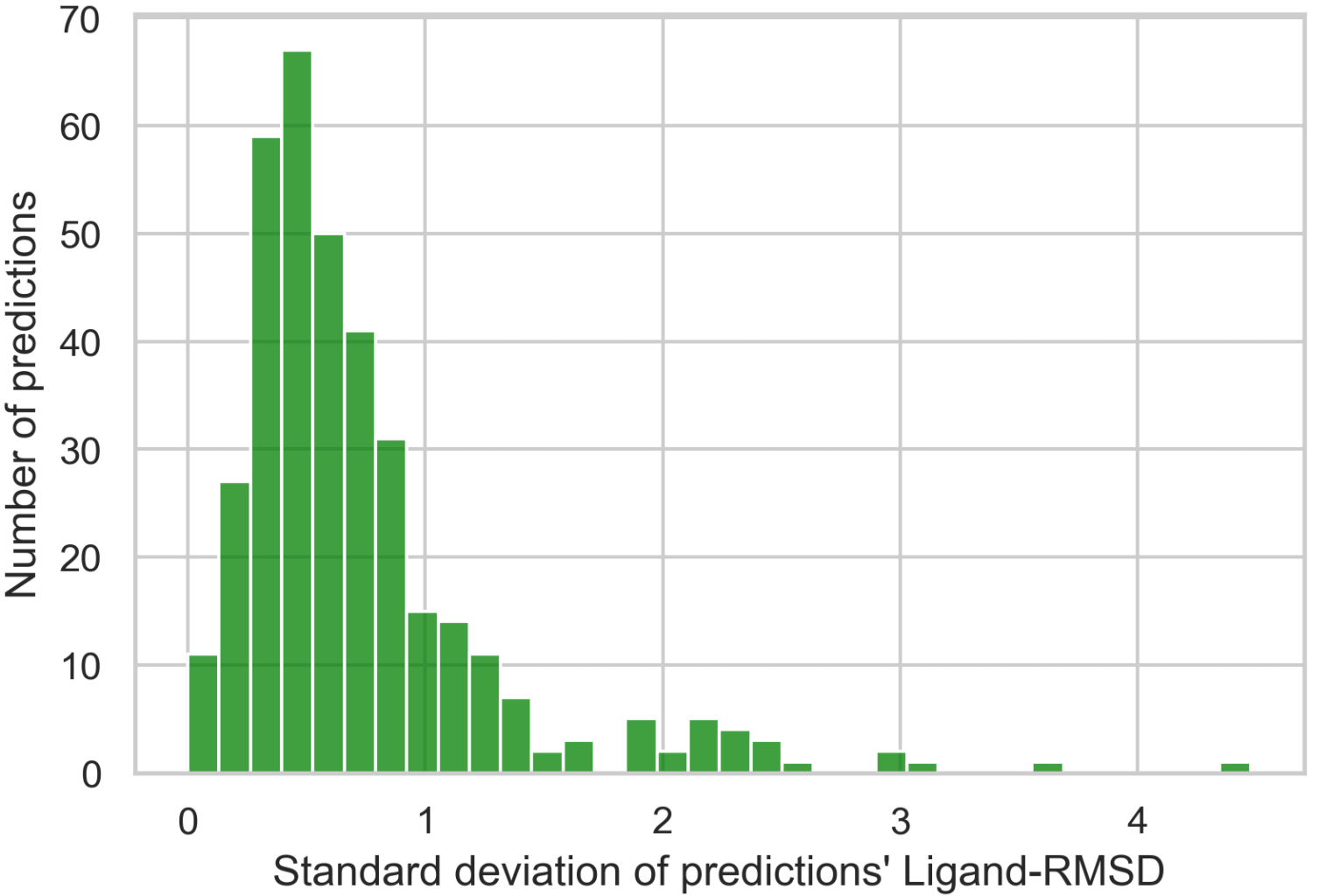}}
\vspace{-0.1cm}
\caption{Histogram of the standard deviations of the L-RMSDs of \textsc{EquiBind}'s predictions when using different initial conformers.}
\label{fig:std-histogram}
\end{center}
 \vskip -0.5cm
\end{figure}

\textbf{Sensitivity to initial conformer.}
\textsc{EquiBind}'s predictions depend on the initial conformer's torsion angles, bond angles, and bond lengths (the baselines only depend on initial bond angles and lengths). In \cref{fig:std-histogram} we investigate the risk of an "unlucky" initial conformer leading to a high L-RMSD. For $363$ complexes, we generate $30$ different initial RDKit conformers. \textsc{EquiBind} predicts a binding structure using each of them, and we obtain 30 L-RMSD values of which we calculate the standard deviation. 
We find a low sensitivity to the initial conformer, with the majority of predictions having a smaller standard deviation than $0.5$

\begin{figure}[htb]
\vspace{-0.cm}
\begin{center}
\centerline{\includegraphics[width=.9\columnwidth]{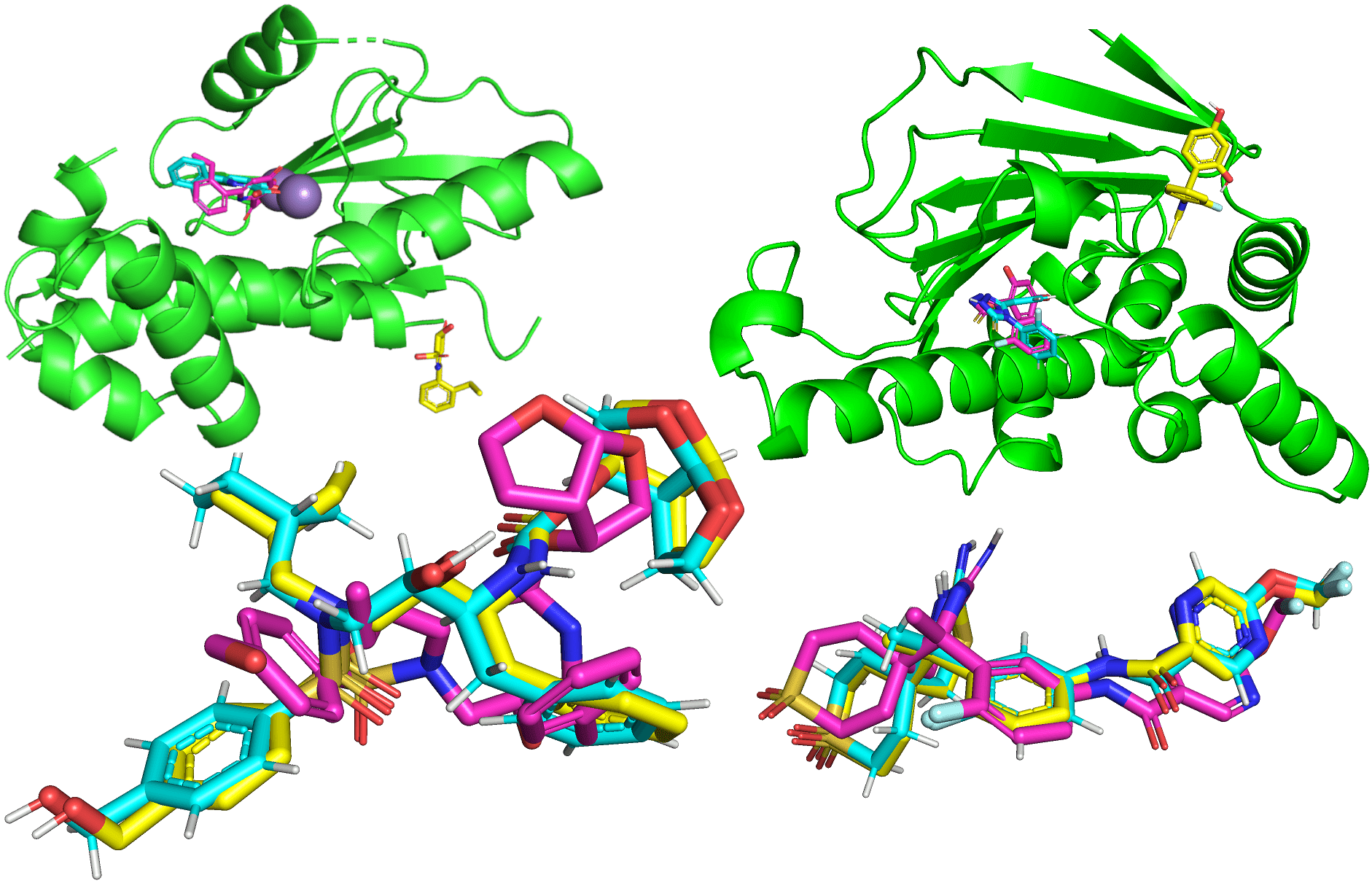}}
\vspace{-0.3cm}
\caption{Two cherry-picked example predictions where \textsc{EquiBind} has better RMSD than GNINA (top) and two where \textsc{EquiBind} performs worse than GNINA (bottom). The ground truth ligand is in cyan, \textsc{EquiBind} in pink, and GNINA in yellow.}
\label{fig:cherry-picked}
\end{center}
 \vskip -0.5cm
\end{figure}

\textbf{Visualizations.}
\textsc{EquiBind}'s predictions are rarely far off from the true ligand, but there are cases where it struggles to find the exact torsion angles and, therefore, the right atom configurations in the ligand. Examples of this are in \cref{fig:cherry-picked} and show two cases where GNINA performs worse and produces a prediction that is far off while \textsc{EquiBind} is able to find the binding location. The other two cases, where GNINA is better, display how the baseline more exactly finds the true structure, but \textsc{EquiBind} still finds the correct approximate location. Further visualizations of predictions are in Appendix Figure \ref{fig:cherry-picked-side-chain}.

\begin{figure}[htb]
\vspace{-0.cm}
\begin{center}
\centerline{\includegraphics[width=0.8\columnwidth]{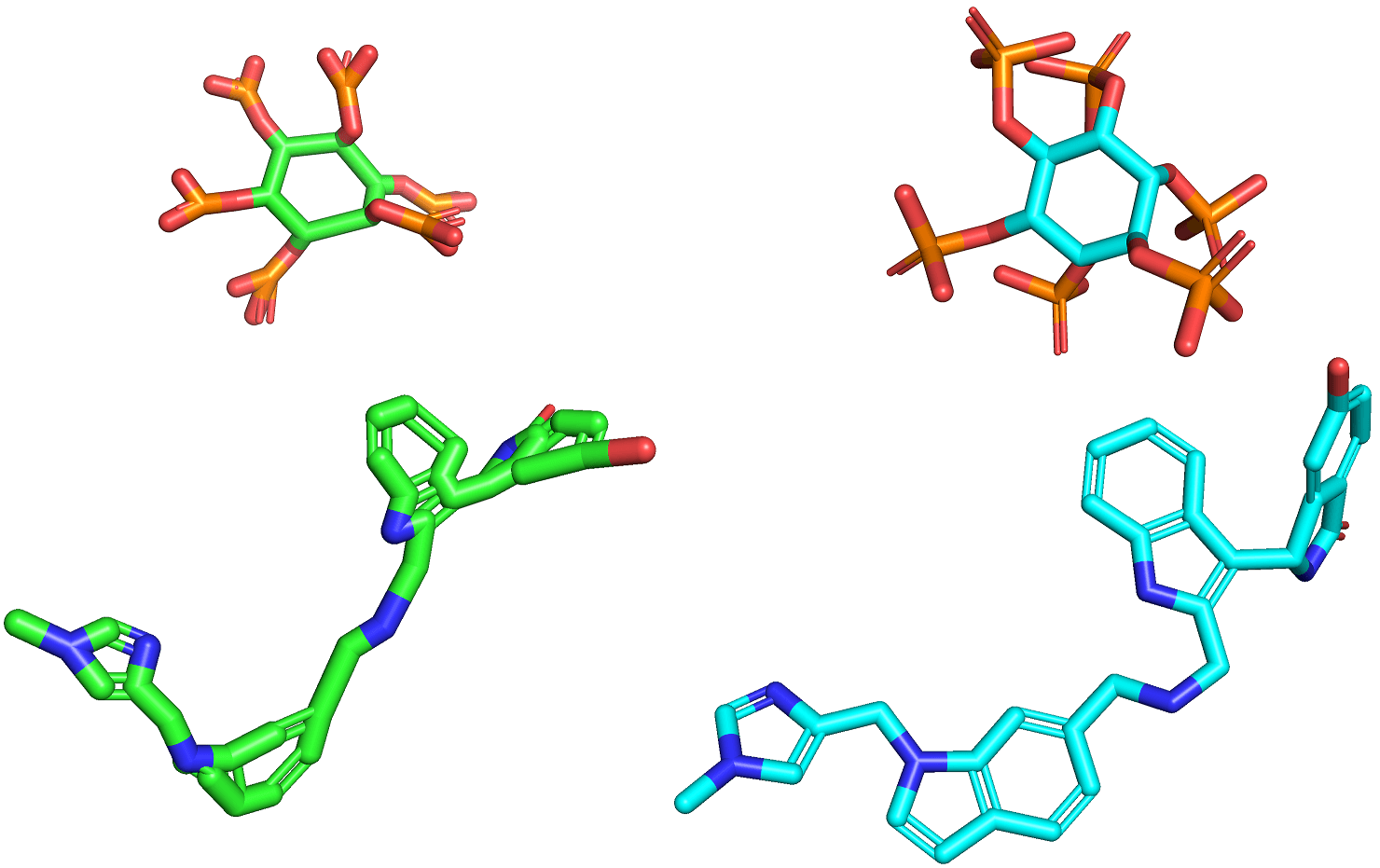}}
\vspace{-0.3cm}
\caption{Left: in green are two uncorrected pseudo-molecules predicted by \textsc{EquiBind}-U. Right: the final output of \textsc{EquiBind} with corrections using our fast flexible conformer fitting applied to produce a conformer with realistic bond angles and lengths. }
\label{fig:conformer-fitting}
\end{center}
 \vskip -0.5cm
\end{figure}

\textbf{Fast point cloud fitting.} In \cref{fig:conformer-fitting} we visualize our novel fast point cloud ligand fitting described in \cref{sssec:conformer-fitting}. The point clouds produced by the uncorrected flexible \textsc{EquiBind}-U are not realistic molecules. The corrections use a conformer with valid bond lengths and angles and change its torsions to most closely match the point cloud.

\textbf{Limitations.}
One drawback of \textsc{EquiBind} is that it only implicitly models the atom positions of side chains. This is done via the local frame encoding features of \citet{Jumper2021} that we employ in the $\alpha$-carbon graph of the receptor. Explicitly representing these atoms might improve precise docking. We experimented with surface atoms and fine-tuning approaches that use an atom subgraph of the receptor with results in Appendix \ref{appendix:results}. However, this yielded only small or no improvements while adding considerable computational complexity. We leave further exploration of this strategy for future work. These results are in line with prior protein modeling techniques such as AlphaFold2, \cite{Jumper2021} which successfully predicts side chains based on only residue level information.

\section{Conclusion}

The promising ability of deep neural networks to predict protein structures has sparked a large amount of research in computational drug discovery. Here, we proposed \textsc{EquiBind}, a deep neural model which relies on SE(3)-equivariant graph neural networks to predict bound protein-ligand conformations in a single shot. Our model shows strong empirical performance against state-of-the-art baselines, and we demonstrate its potential in a hybrid workflow by combining it with existing fine-tuning methods. We expect that  \textsc{EquiBind} and similar models will progress the adoption of deep learning in drug discovery.

\section{Acknowledgments}
We dedicate this paper to the late first author Octavian-Eugen Ganea (1987-2022), a brilliant researcher and a generous mentor. Over the course of his career Octavian already made multiple significant contributions to geometric ML and molecular modeling. He had a promising, bright future of stellar discoveries ahead of him. 

The authors thank Bruno Correia, Gabriele Corso, Simon Batzner, Miguel García-Ortegón, Vignesh Ram Somnath, Stiefl Nikolaus, Arne Schneuing, Julia Buhmann, David Ryan Koes, Patrick Walters, Michael Heinzinger, Tian Xie, Xiang Fu, and Jeremy Wohlwend for insightful discussions and valuable feedback.

Octavian Ganea was funded by the Machine Learning for Pharmaceutical Discovery and Synthesis (MLPDS) consortium, the Abdul Latif Jameel Clinic for Machine Learning in Health, the DTRA Discovery of Medical Countermeasures Against New and Emerging (DOMANE) threats program, and the DARPA Accelerated Molecular Discovery program.
Lagnajit Pattanaik is funded by the MLPDS consortium and the MIT-Takeda Fellowship. Regina Barzilay and Tommi Jaakkola also acknowledge support from NSF Expeditions grant (award 1918839): Collaborative Research: Understanding the World Through Code.

\bibliography{references}
\bibliographystyle{icml2022}

\newpage
\appendix
\onecolumn

\section{Additional Results}\label{appendix:results}


\begin{figure}[htb]
\vspace{-0.cm}
  \centering
\includegraphics[width=0.47\textwidth]{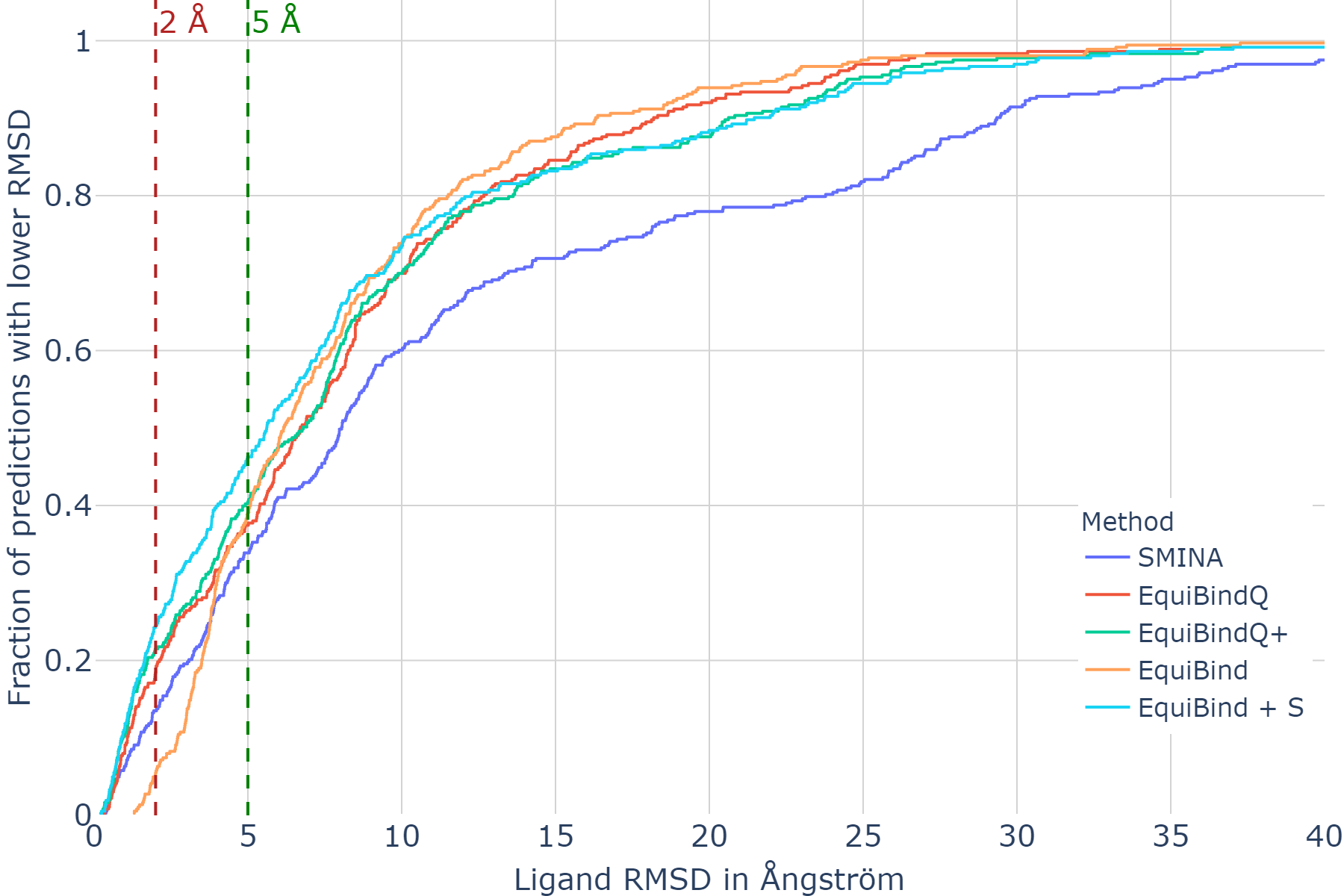}
  \hfill
\includegraphics[width=0.47\textwidth]{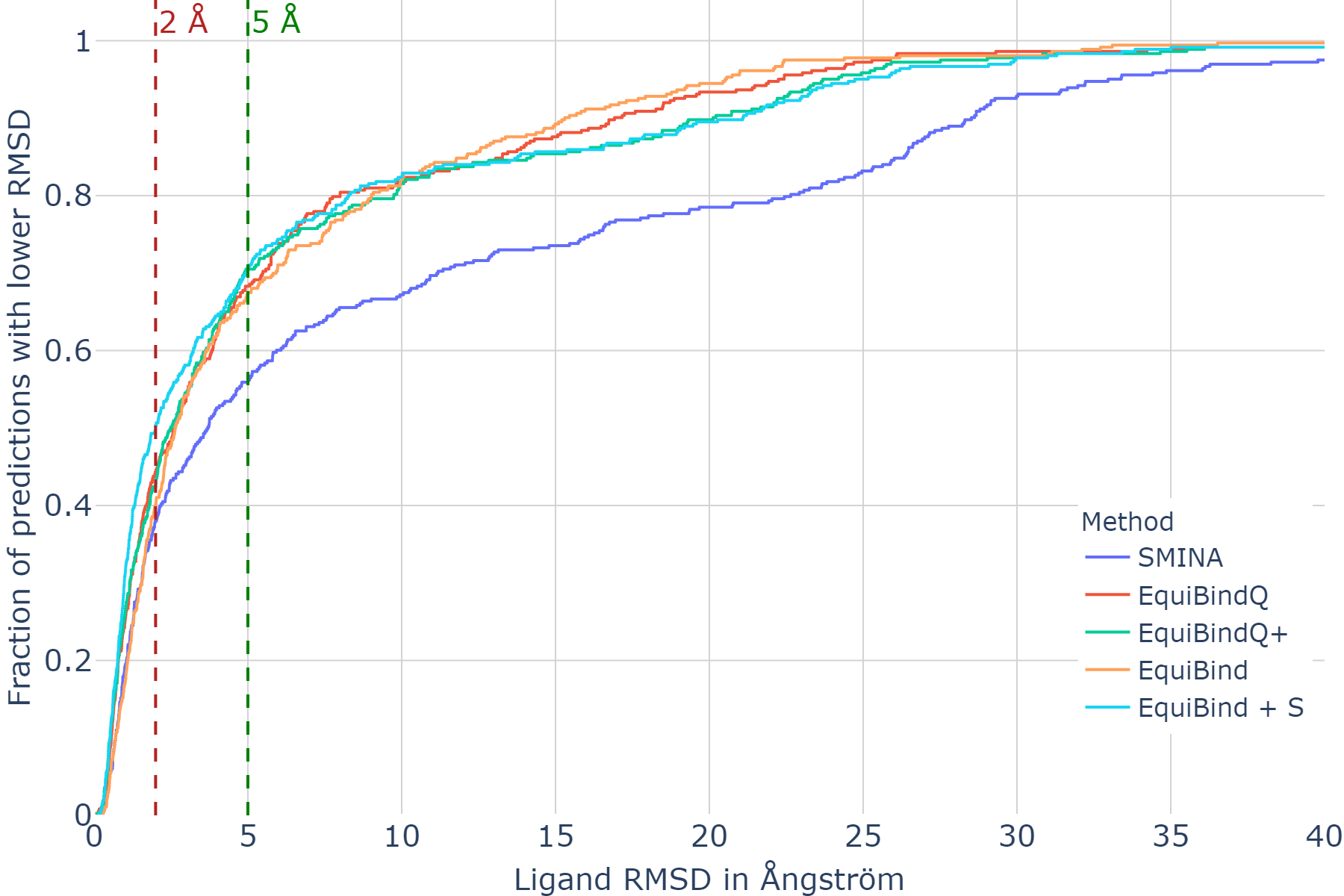}
\vspace{-0.cm}
  \caption{\textbf{Flexible blind self-docking.} Left: Cumulative histogram of the L-RMSD. Right: Cumulative histogram of the centroid distance.} \label{fig:additional-histogram}
\end{figure}

\begin{figure}[htb]
\vspace{-0.cm}
  \centering
\includegraphics[width=0.47\textwidth]{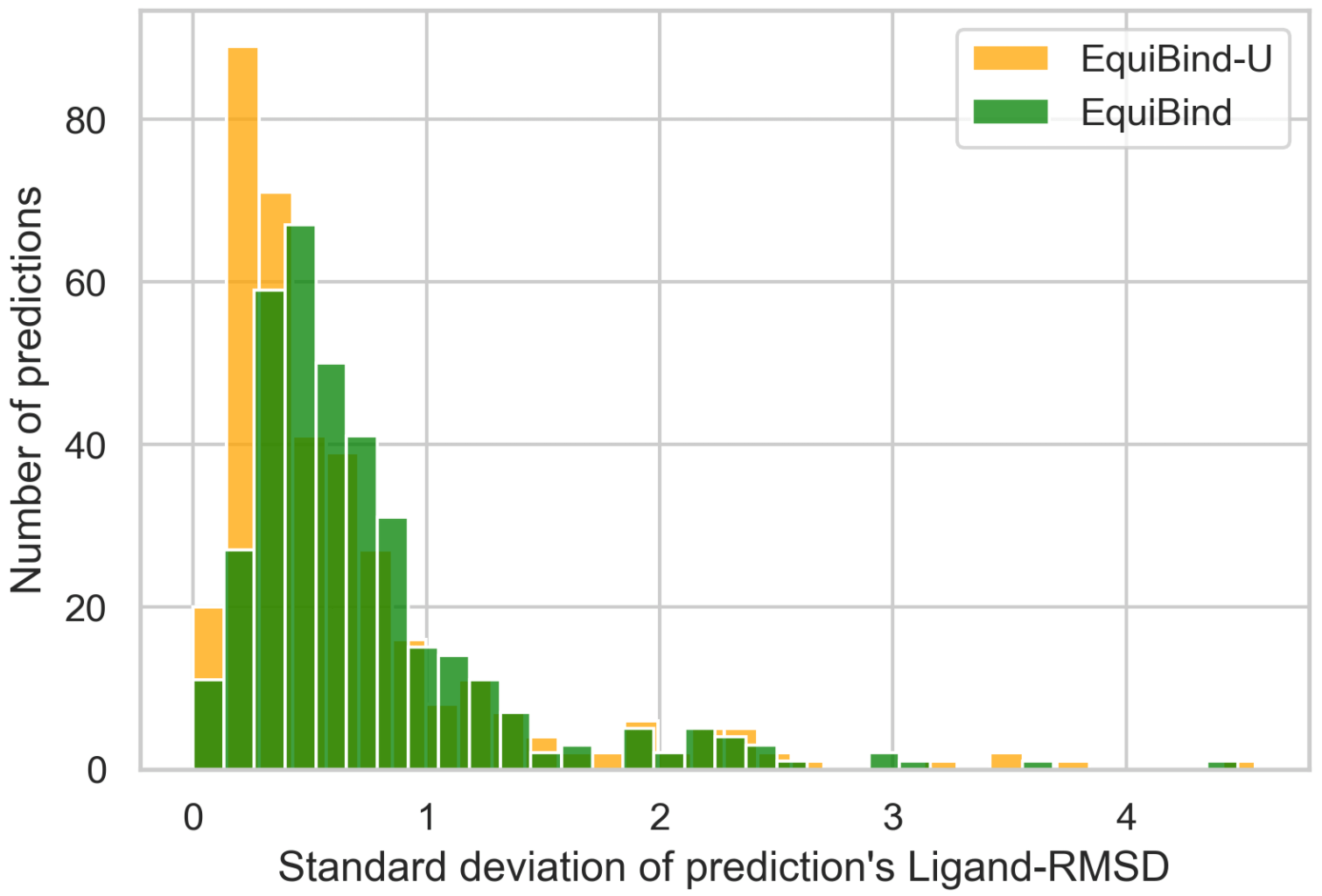}
  \hfill
\includegraphics[width=0.47\textwidth]{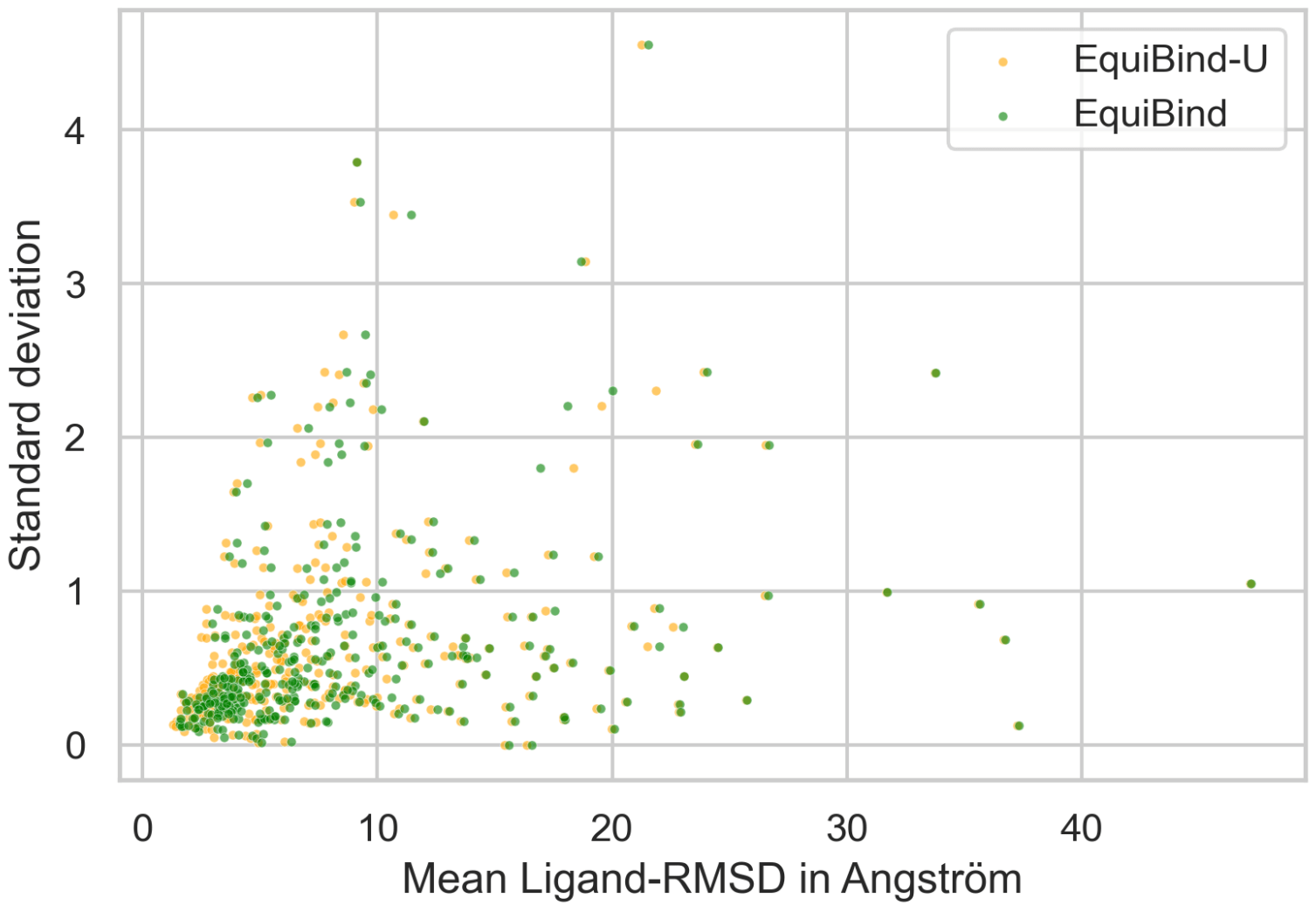}
\vspace{-0.cm}
  \caption{\textbf{Sensitivity to initial conformer.} Left: Histogram of the standard deviations of the L-RMSDs of \textsc{EquiBind}'s and \textsc{EquiBind}-U's predictions when using 30 different initial conformers. Right: Scatter-plot with a point for each complex showing the mean L-RMSD and standard deviation of \textsc{EquiBind}'s or \textsc{EquiBind}-U's 30 predictions from 30 different initial input conformers.} \label{fig:std-histogram-and-scatterplot}
\end{figure}

\begin{table}[htb]
    
    \caption{\textbf{Flexible blind self-docking.} All methods receive a random RDKit conformer of the ligand as input and are tasked to find the binding site and the correct binding structure. Comparison of \textsc{EquiBind}-SA which additionally uses surface atoms of the receptor and \textsc{EquiBind}-A, which makes a prediction using \textsc{EquiBind}-R and refines it using an all-atom subgraph in a 10 \AA{} radius around the predicted ligand. The methods have to be compared with \textsc{EquiBind}-U as corrections would still need to be applied. Additionally shown is \textsc{EquiBind w/o DG} which refers to \textsc{EquiBind} without the LAS-DG regularization.}
    \label{tab:results_3}
     \vspace{-0.3cm}
     \centering
    \begin{small}
     \begin{center}
     \begin{sc}
    \begin{tabular}{lcc|cccccc|cccccc|cc}
    \toprule
     & & &\multicolumn{6}{c}{Ligand RMSD $\downarrow$} & \multicolumn{6}{c}{Centroid Distance $\downarrow$} & \multicolumn{2}{c}{Kabsch}\\
     &\begin{tabular}{@{}c@{}}\ avg.\\sec.\end{tabular} & \begin{tabular}{@{}c@{}}\ avg.\\sec.\end{tabular} &\multicolumn{4}{c}{Percentiles $\downarrow$} & \multicolumn{2}{c}{\begin{tabular}{@{}c@{}}\% below\\threshold $\uparrow$\end{tabular} }  &\multicolumn{4}{c}{Percentiles $\downarrow$} & \multicolumn{2}{c}{\begin{tabular}{@{}c@{}}\% below\\thresh. $\uparrow$\end{tabular} } & \multicolumn{2}{c}{RMSD $\downarrow$}\\
    
    \textbf{Methods} & 16-CPU & GPU & Mean & 25th & 50th & 75th & 5 \AA{}  &  2 \AA{} & Mean & 25th & 50th & 75th & 5 \AA{}  &  2 \AA{} & Mean & Med.\\
    \midrule
     \textsc{EquiBind}-SA &0.14& 0.02& 8.6 & 3.6 & 6.1 &  11.4 & 41.3 & 3.8 & 6.1 & 1.4  & 2.7 & 7.3 & 66.6&  40.5 & 2.4 & 2.0\\
     \textsc{EquiBind}-A &5.56& 5.44 & 8.2 & 3.8 & 6.0 &  10.2 & 43.0 & 3.3 & 6.1 & 1.7 & 3.3 & 7.0 & 63.3&  30.3 & 3.0 & 2.7\\
    \textsc{EquiBind}-U &0.14& 0.02 & 7.8 & 3.3 & 5.7 &  9.7 & 42.4 & 7.2 & 5.6 & 1.3 & 2.6 & 7.4 & 67.5&  40.0 & 2.1 & 1.8\\
    \textsc{EquiBind w/o DG} &0.16& 0.04 & 8.7 & 3.9 & 6.1 &  11.3 & 39.1 & 4.1 & 6.1 & 1.3 & 2.7 & 7.4 & 67.6&  39.3 & 2.6 & 2.3\\
     \bottomrule
    \end{tabular}
    \end{sc}
    \end{center}
    \end{small}
    \vspace{-0.3cm}
\end{table}

In \cref{tab:results_3} we compare \textsc{EquiBind} which only implicitly uses atom level locations with approaches that explicitly use atoms as nodes in the processed graph. \textsc{EquiBind}-SA uses the surface atoms of the receptor, which are found using the MSMS tool. This additional step does not impact inference runtime, which we measure without preprocessing as explained in Section \ref{eval_setup}. \textsc{EquiBind}-A makes a prediction using \textsc{EquiBind}-R and uses an all-atom subgraph in a 10 \AA{} radius around the predicted ligand to further refine the prediction. This additional step significantly impacts inference runtime. Both methods require around 2-3 times more GPU RAM than \textsc{EquiBind}.

\begin{figure}[htb]
\vspace{-0.cm}
  \centering
\includegraphics[width=0.47\textwidth]{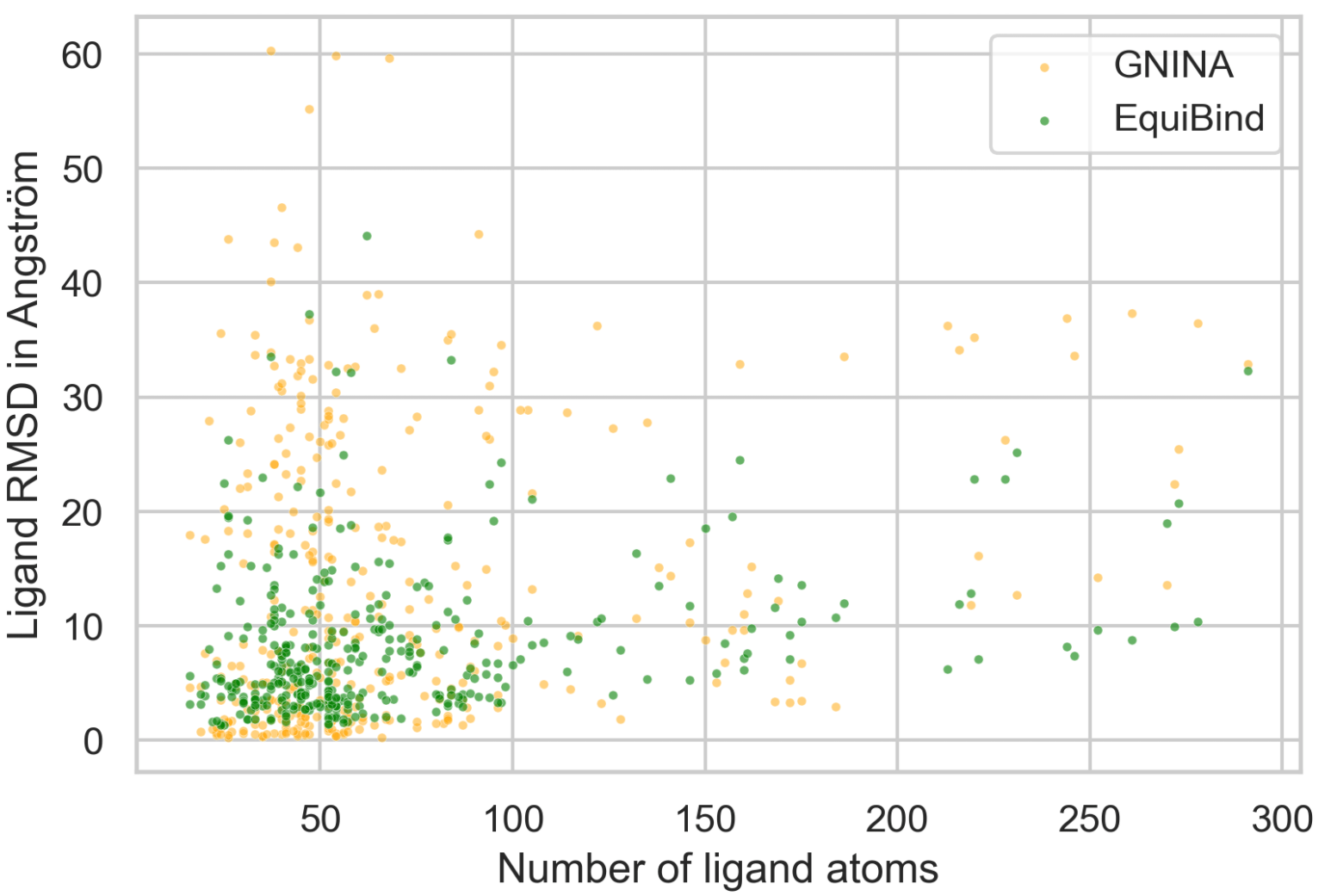}
  \hfill
\includegraphics[width=0.47\textwidth]{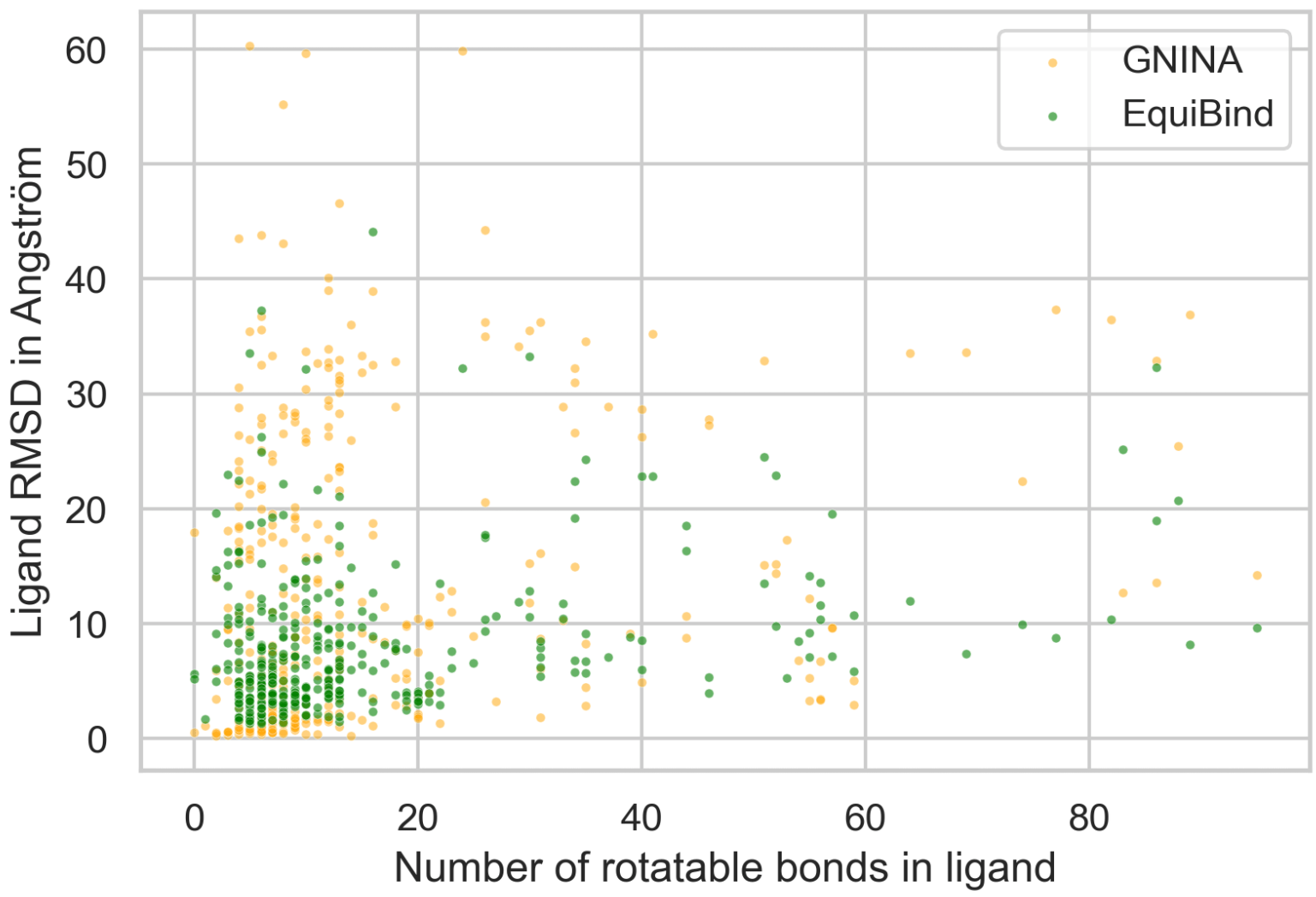}
\vspace{-0.cm}
  \caption{\textbf{Performance correlation with molecule size.} Left: Scatter-plot showing the L-RMSD and the number of ligand atoms for each prediction of \textsc{EquiBind} and GNINA. Right: The same for the number of rotatable bonds in the ligand.} \label{fig:rmsd_vs_ligand_size_scatter_GNINA}
\end{figure}

\begin{figure}[htb]
\vspace{-0.cm}
  \centering
\includegraphics[width=0.47\textwidth]{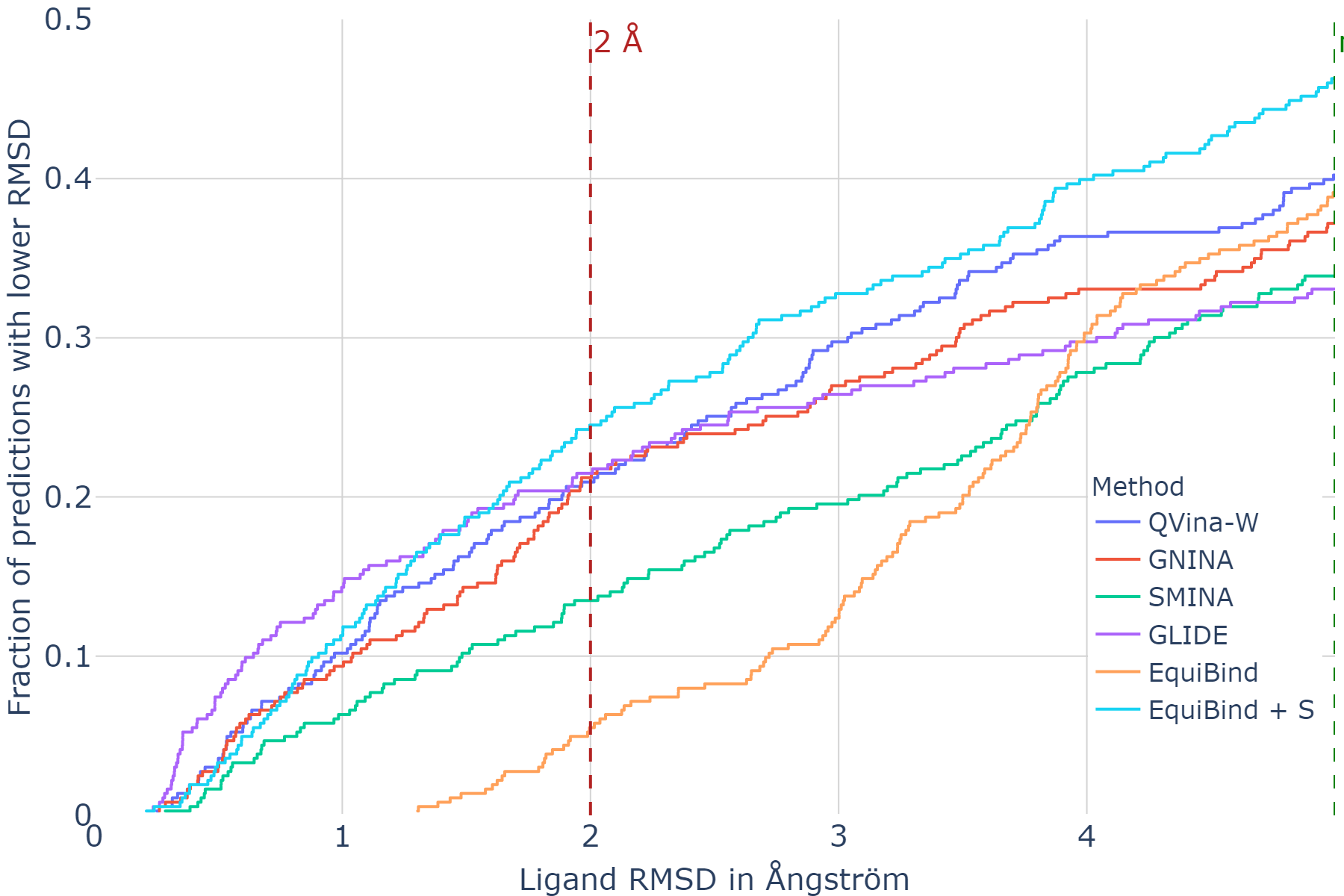}
  \hfill
\includegraphics[width=0.47\textwidth]{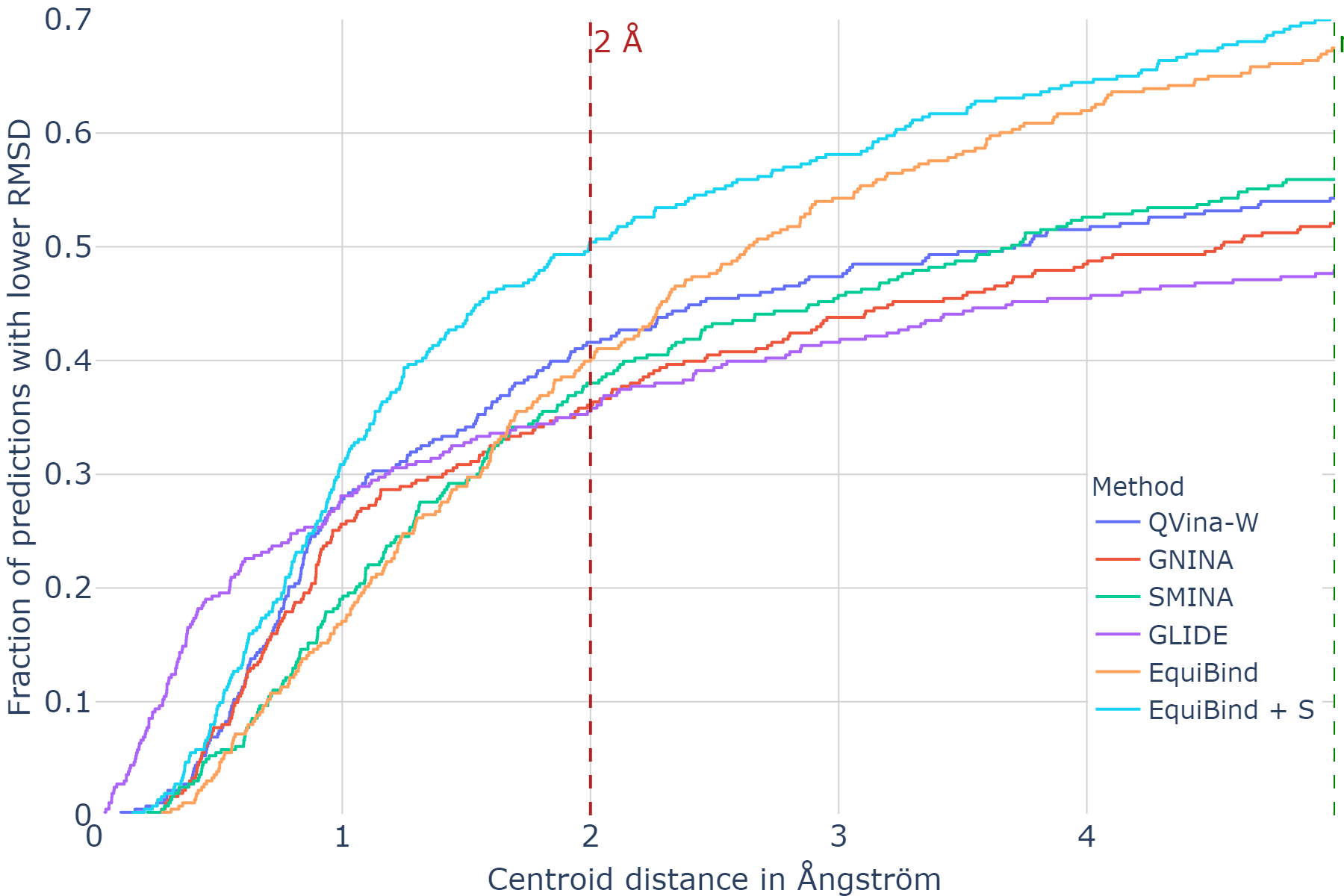}
\vspace{-0.cm}
  \caption{\textbf{Flexible blind self-docking.} Zoomed in versions of the histograms in Figure \ref{fig:histograms} where only the 0-5 \AA{} range is shown. Left: Cumulative histogram of the L-RMSD. Right: Cumulative histogram of the centroid distance.} \label{fig:histogram-zoomed-in}
\end{figure}

In \cref{fig:histogram-zoomed-in} we can again observe that \textsc{EquiBind} struggles to produce many predictions in the low L-RMSD range. When adding an additional fine-tuning step such as with \textsc{EquiBindS}, the model is able to match or outperform the baselines in all L-RMSD ranges. Thus predictions speed can be traded off for additional accuracy via fine-tuning \textsc{EquiBind}'s predictions with classical physics-based methods. Above the 3.8 \AA{} cutoff, the vanilla \textsc{EquiBind} starts outperforming the baselines even without fine-tuning. 

\begin{figure}[ht]
\vspace{-0.cm}
\begin{center}
\centerline{\includegraphics[width=\columnwidth]{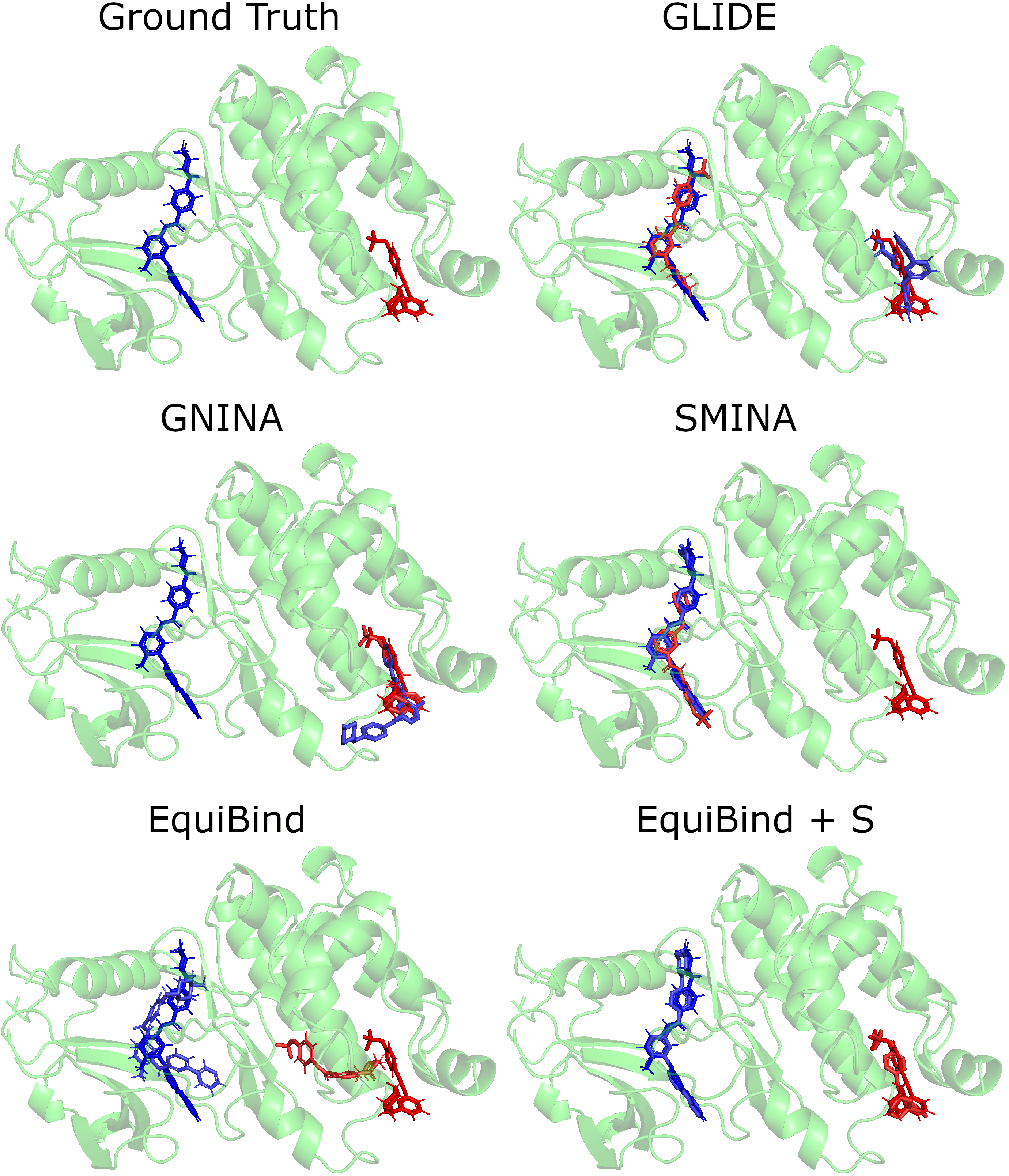}}
\vspace{-0cm}
\caption{Case study for the Tyrosine Kinase 6HD6 (in green) and two inhibitor drugs (in red and blue) for lung cancer, leukemia, and gastrointestinal tumors: ABL001 (Asciminib - Novartis, FDA approved Oct 2021) and Imatinib (STI-571 - Novartis). Shown is always the ground truth for both ligands and the predictions of the models in the same color, only lighter. Neither the ligands nor the receptor was part of the training set. We find that only \textsc{EquiBind} and \textsc{EquiBind + S} are able to identify the correct binding pocket and \textsc{EquiBind + S} almost perfectly predicts the ligand structures. This complex was not cherry-picked - it was suggested to us by industry collaborators as an important and hard to dock complex.  }
\label{fig:tyrosine-case-study}
\end{center}
 \vskip -0in
\end{figure}

\begin{figure}[htb]
\vspace{-0.cm}
  \centering
\includegraphics[width=0.47\textwidth]{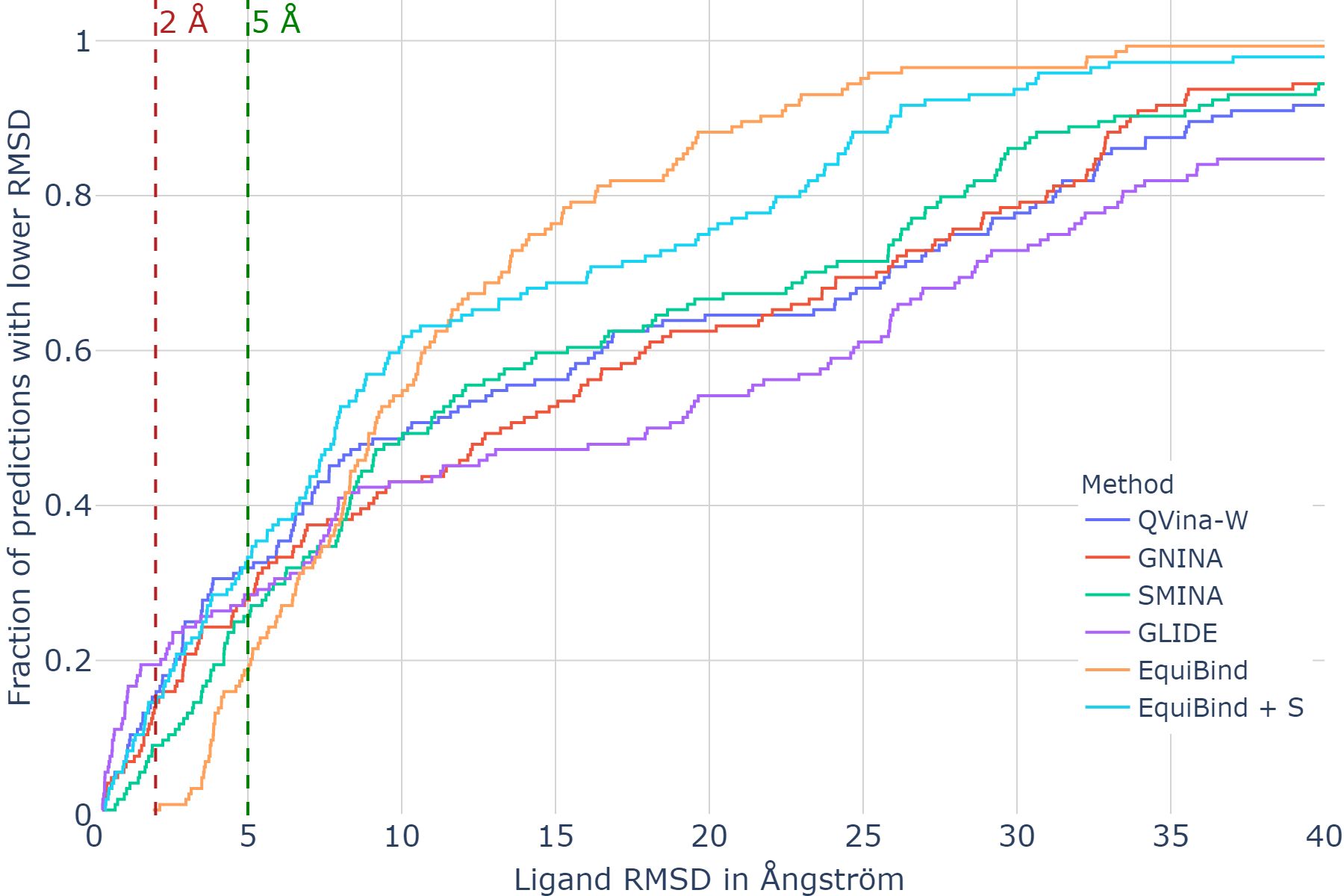}
  \hfill
\includegraphics[width=0.47\textwidth]{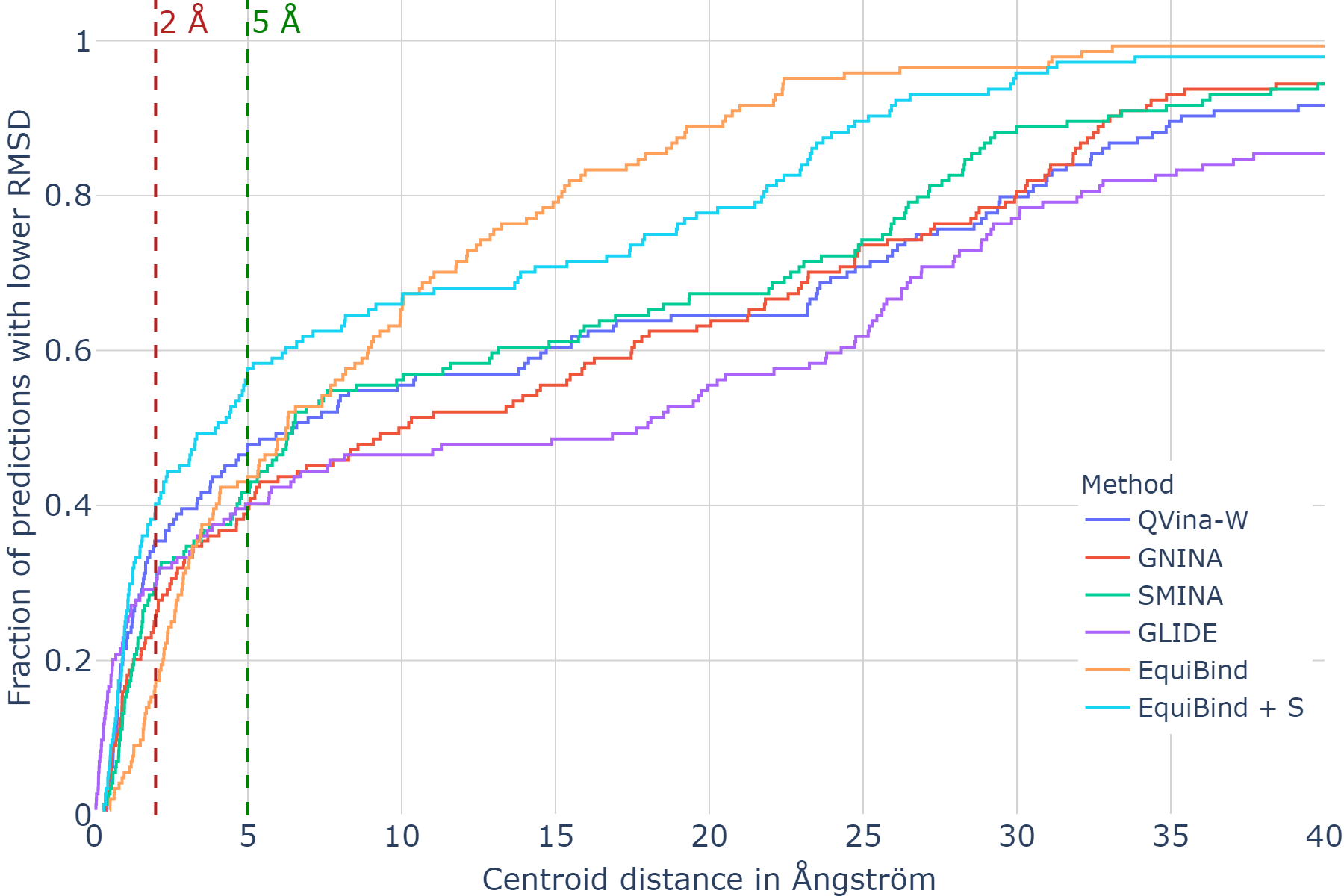}
\vspace{-0.cm}
  \caption{\textbf{Flexible blind self-docking new receptors.} Results when removing all complexes from the time split based test set whose receptor was present in a complex that is older than 2019. Cumulative density histogram of the L-RMSD (top) and centroid distance (bottom) of \textsc{EquiBind} with and without SMINA for fine-tuning.} \label{fig:histogram-new-receptors}
\end{figure}

\setlength{\tabcolsep}{2pt}
\begin{table*}[h!]
    \caption{\textbf{Flexible blind self-docking new receptors.} Results when removing all complexes from the time split based test set whose receptor was present in a complex that is older than 2019. 144 complexes remain. The run times are still averaged over all test complexes. All methods receive a random RDKit conformer of the ligand molecule as input and are tasked to find its binding site and the right orientation + conformation in which it binds. \textsc{EquiBind}-U refers to the model producing uncorrected atomic point clouds $\Zb$ that are not necessarily chemically plausible ligands. \textsc{EquiBind} performs our fast conformer fitting corrections -- see \cref{sssec:conformer-fitting}. \textsc{EquiBind+Q} predicts an approximate ligand position and fine-tunes it using QuickVina 2. \textsc{EquiBind+Q2} samples more candidate positions, and \textsc{EquiBindS} uses SMINA for fine-tuning. *GLIDE runtime details in Appendix \ref{appendix:implementation}.}
    \label{tab:self-docking-noLorR}
     \begin{small}
     \begin{center}
     \begin{sc}
    \begin{tabular}{lcc|cccccc|cccccc|cc}
    \toprule
     & & &\multicolumn{6}{c}{Ligand RMSD $\downarrow$} & \multicolumn{6}{c}{Centroid Distance $\downarrow$} & \multicolumn{2}{c}{Kabsch}\\
     &\begin{tabular}{@{}c@{}}\ avg.\\sec.\end{tabular} & \begin{tabular}{@{}c@{}}\ avg.\\sec.\end{tabular} &\multicolumn{4}{c}{Percentiles $\downarrow$} & \multicolumn{2}{c}{\begin{tabular}{@{}c@{}}\% below\\threshold $\uparrow$\end{tabular} }  &\multicolumn{4}{c}{Percentiles $\downarrow$} & \multicolumn{2}{c}{\begin{tabular}{@{}c@{}}\% below\\thresh. $\uparrow$\end{tabular} } & \multicolumn{2}{c}{RMSD $\downarrow$}\\
    
    \textbf{Methods} & 16-CPU & GPU & Mean & 25th & 50th & 75th & 5 \AA{}  &  2 \AA{} & Mean & 25th & 50th & 75th & 5 \AA{}  &  2 \AA{} & Mean & Med.\\
    \midrule
     \textsc{QVina-W}& 49& -& 16.9 & 3.4  &  10.3 &  28.1 & 31.9 & 15.3 & 15.2  &  1.3 &  6.5 & 26.8 & 47.9 & 35.4 & \textbf{2.2} & 1.9\\
     \textsc{GNINA}& 247& 146& 16.7 & 4.5 & 13.4   & 27.8 & 27.8  & 13.9 & 15.1 & 2.0 &  10.1 & 27.0 & 39.5 & 25.7 & 2.3 & 1.8\\
    \textsc{SMINA}&146& - & 15.7 & 4.8 & 10.9 & 26.0 & 25.7 & 9.0 & 13.6 & 1.6 &  6.5 & 25.7 & 41.7 & 29.9 & 2.3 & 1.9\\
    \textsc{GLIDE}& 1405*& - & 19.6 & \textbf{3.4} & 18.0   & 31.4 & 28.7 & \textbf{19.6} & 18.1& 1.1 &  17.6 & 29.1 & 40.6 & 29.4 & 2.3 & \textbf{1.7}\\
    \hline
    \textsc{EquiBind} &\textbf{0.16}& \textbf{0.04} & \textbf{11.3} & 5.9 & 9.1 &  \textbf{14.3} & 18.8 & 0.7 & \textbf{8.9} & 2.6 & 6.3 & \textbf{12.9} & 43.8 &  16.7 & 2.7 & 2.2\\
     \textsc{EquiBind+Q}& 8& 8 & 11.5 & 5.5 & 8.7 &  15.7 & 22.9 & 9.3 &  9.1 & 1.6 & 5.5 &  14.0 & 47.9 & 30.7  &2.3 & 1.9\\
     \textsc{EquiBind+Q2}& 15& 15 & 12.0 & 4.1 & 8.0 &  19.7 & 28.0 & 13.3 &  9.8 & 1.3  & 4.4 &  18.5 & 53.9 & 30.8  & 2.3 & \textbf{1.7}\\
     \textsc{EquiBind+S}& 146& 146 & 11.9 & 3.6 & \textbf{7.9} &  19.7 & \textbf{33.3} & 14.6 &  9.7 & \textbf{1.0}  & \textbf{4.0} &  18.2 & \textbf{57.6} & \textbf{40.3}  & \textbf{2.2}& 1.8\\
     \midrule
     \textsc{EquiBind}-U &0.14& 0.02 & 11.0 & 5.7 & 8.8 &  14.1 & 21.5 & 1.4 & 8.9 & 2.6 & 6.3 & 12.9 & 43.8 &  16.7 & 2.2 & 1.8\\
     \bottomrule
    \end{tabular}
    \end{sc}
    \end{center}
    \end{small}
     \vskip -0.25in
\end{table*}

\begin{table}[htb]
    
    \caption{Standard deviations averaged over 10 training runs with different weight initialization for \textsc{EquiBind} in Table \ref{tab:self-docking}}
    \label{tab:stdevs}
     \centering
    \begin{small}
     \begin{center}
     \begin{sc}
    \begin{tabular}{lcc|cccccc|cccccc|cc}
    \toprule
     & & &\multicolumn{6}{c}{Ligand RMSD $\downarrow$} & \multicolumn{6}{c}{Centroid Distance $\downarrow$} & \multicolumn{2}{c}{Kabsch}\\
    
    \textbf{Methods} &  &  & Mean & 25th & 50th & 75th & 5 \AA{}  &  2 \AA{} & Mean & 25th & 50th & 75th & 5 \AA{}  &  2 \AA{} & Mean & Med.\\
    \midrule
    \textsc{EquiBind}-U & &  &  0.23 & 0.07 & 0.25 & 0.39 & 2.25 & 0.98 & 0.28 & 0.05 & 0.14 & 0.49 & 1.71 & 1.75 & 0.02 & 0.04 \\

     \bottomrule
    \end{tabular}
    \end{sc}
    \end{center}
    \end{small}
    \vspace{-0.5cm}
\end{table}

\begin{figure}[htb]
\vspace{-0.cm}
  \centering
\includegraphics[width=0.7\textwidth]{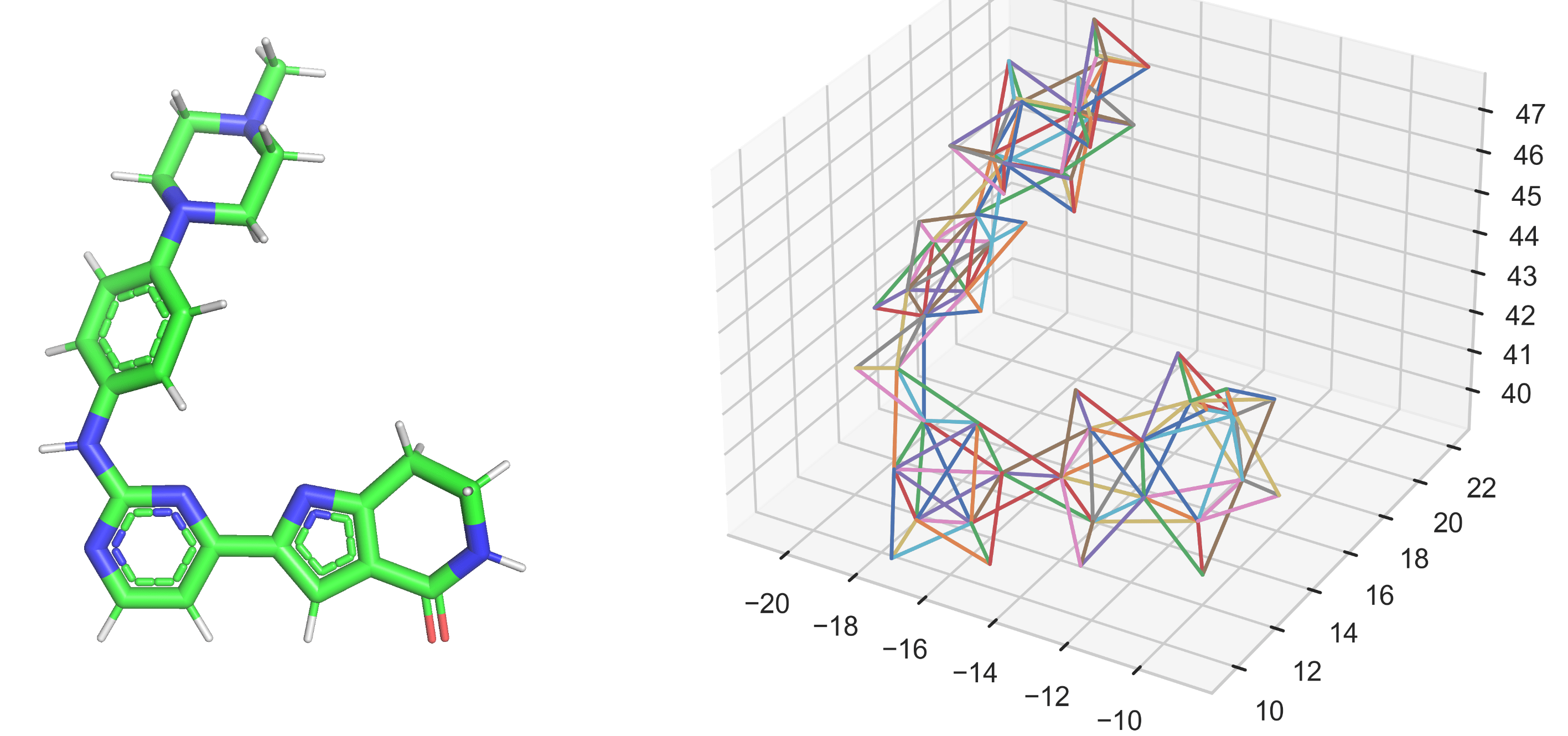}
\vspace{-0.cm}
  \caption{\textbf{LAS distance geometry constraint visualization.} Visualization of the interatomic distances which are included in the LAS distance geometry constraints in Equation \ref{equation:dg}. The pairwise distances in rings are only included if the ring is aromatic, like in the bottom left ring of the depicted molecule. The torsion angles in non-aromatic rings, such as the one in the bottom right, remain flexible. } \label{fig:dg-visualization}
\end{figure}

\begin{figure}[ht]
\vspace{-0.cm}
\begin{center}
\centerline{\includegraphics[width=\columnwidth]{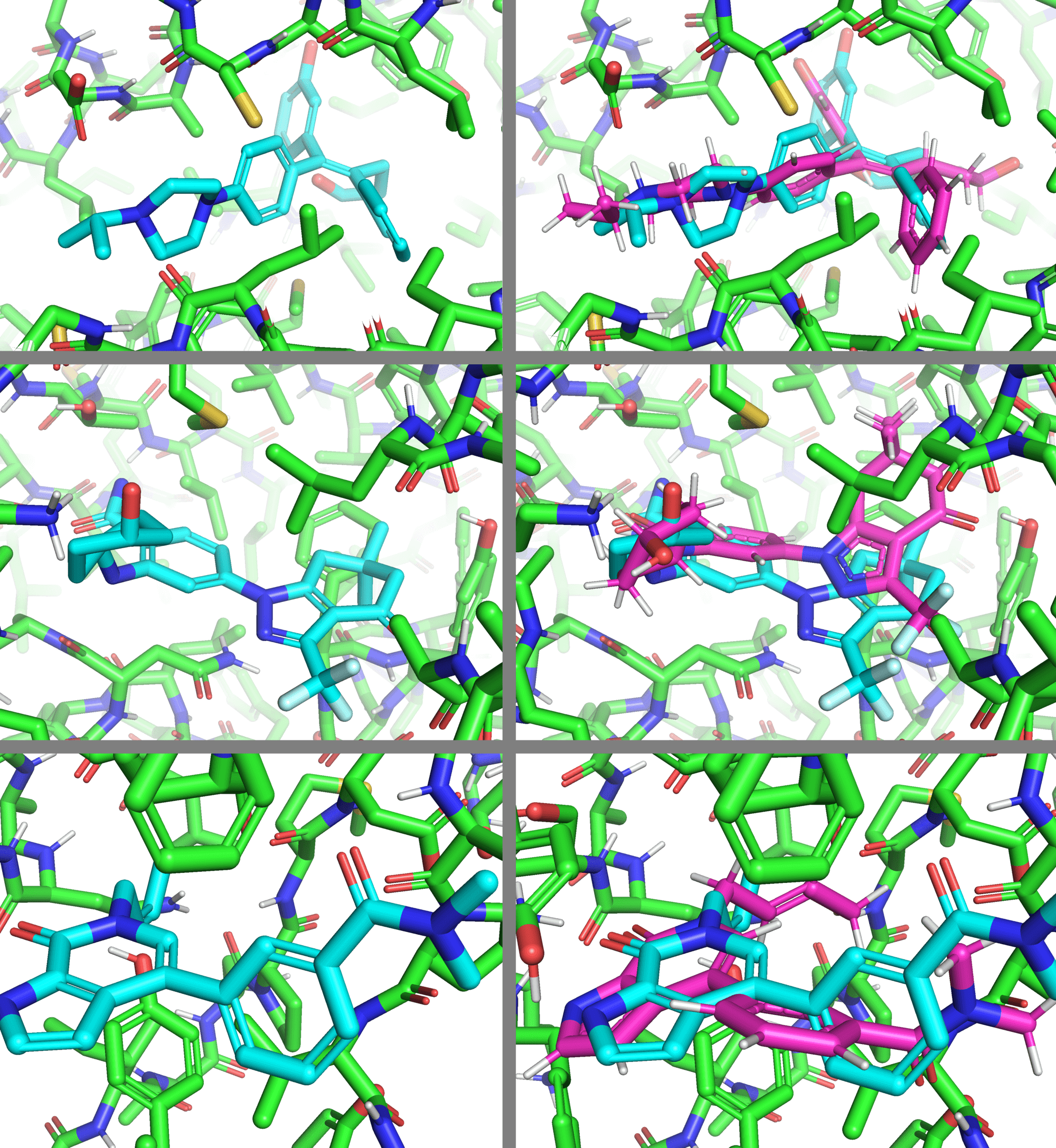}}
\vspace{-0cm}
\caption{Three cherry-picked example predictions of \textsc{EquiBind} (cyan) on the left with the same image on the right, including the true bound conformer in pink.}
\label{fig:cherry-picked-side-chain}
\end{center}
 \vskip -0in
\end{figure}

\clearpage

\section{Dataset}\label{appendix:dataset}
\textbf{Preprocessing.} The time split is done after preprocessing the $19\,443$ complexes of PDBBind v2020 as follows. First, we drop all complexes that cannot be processed 
by the RDKit library~\citep{rdkit}, leaving $19\,119$ complexes. We process each ligand and receptor with OpenBabel \cite{openbabel}. Next we correct all receptor hydrogens and add missing ones using \texttt{reduce}\footnote{\url{https://github.com/rlabduke/reduce}}. 

\begin{figure}[htbp]

  \centering
 
\includegraphics[width=0.47\textwidth]{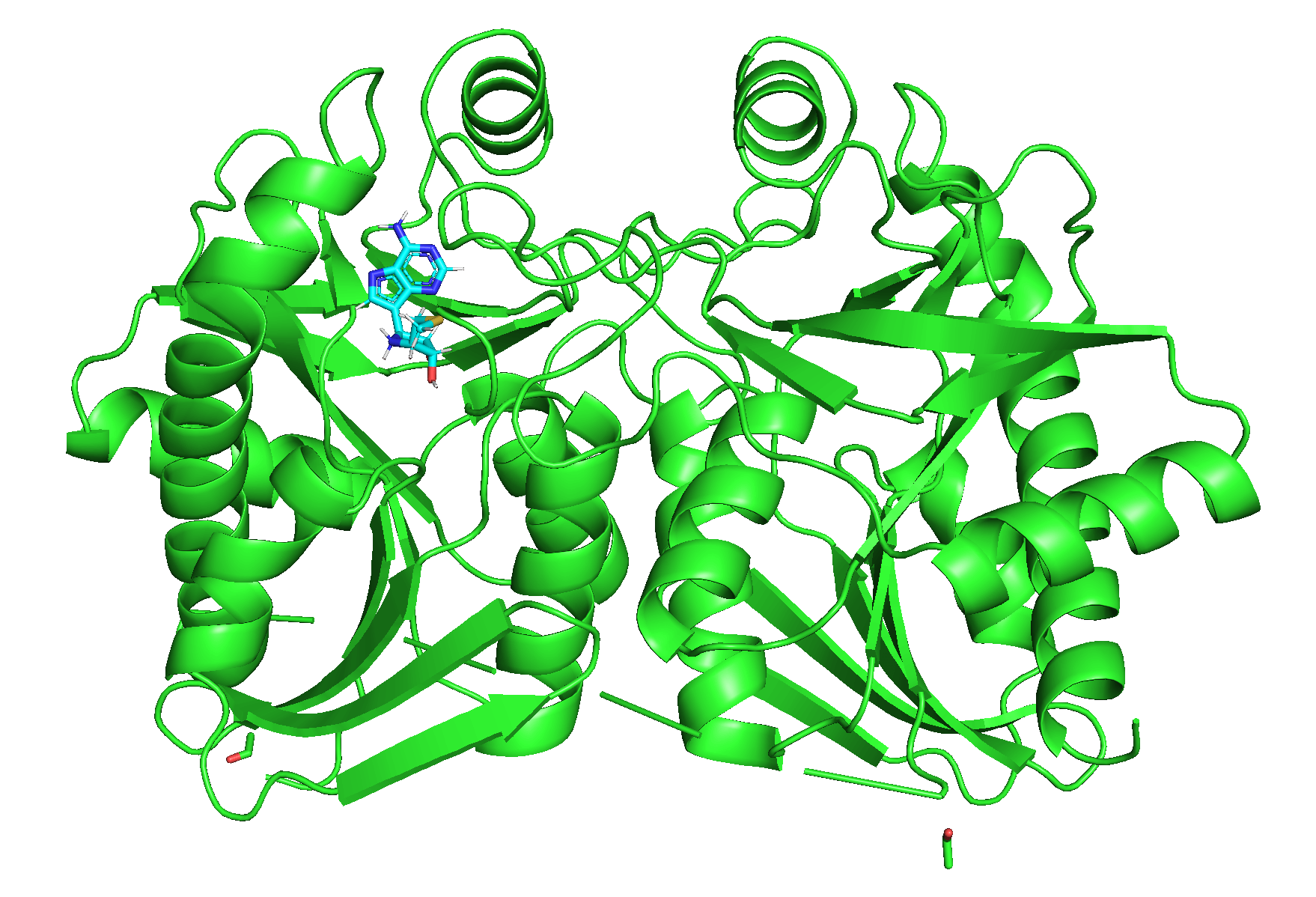}
  \hfill
\includegraphics[width=0.47\textwidth]{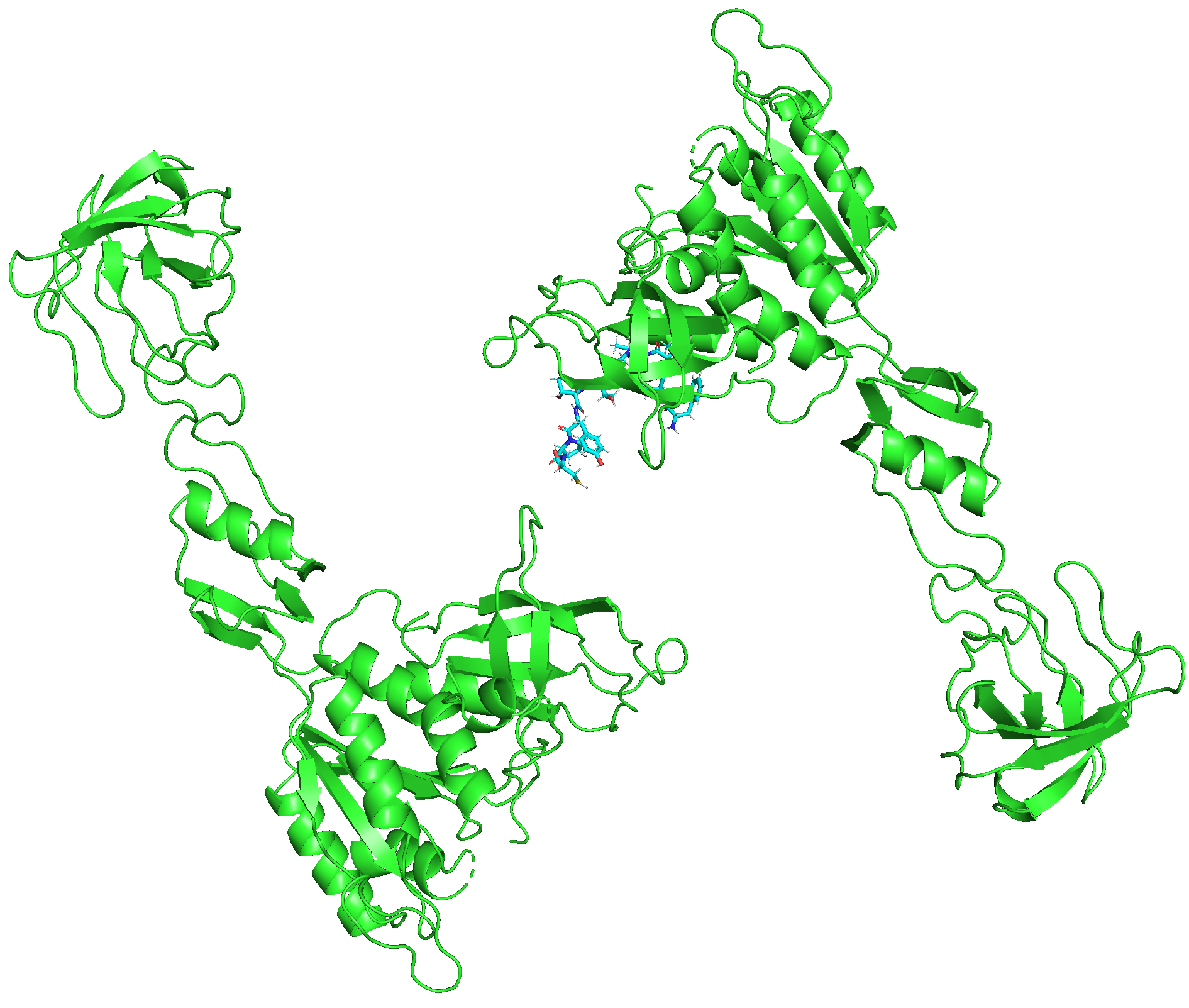}
  \caption{Examples of symmetric receptor complexes with multiple equally valid binding positions for the ligand.} \label{symmetric-receptors}

\end{figure}

A significant remaining data issue is symmetric receptor structures comprised of the same protein repeated multiple times. In these cases, the ligand could equally likely bind to the pocket of each of the proteins, i.e., multiple ligand correct positions are possible. However, the ground truth ligand is only placed in one of those locations. Examples are in  \cref{symmetric-receptors}. We address the majority of these cases by only keeping the connected components of the receptor, which have an atom within a 10 \AA{}\ radius of any ligand atom.

\begin{figure}[htbp]

  \centering
 
\includegraphics[width=0.47\textwidth]{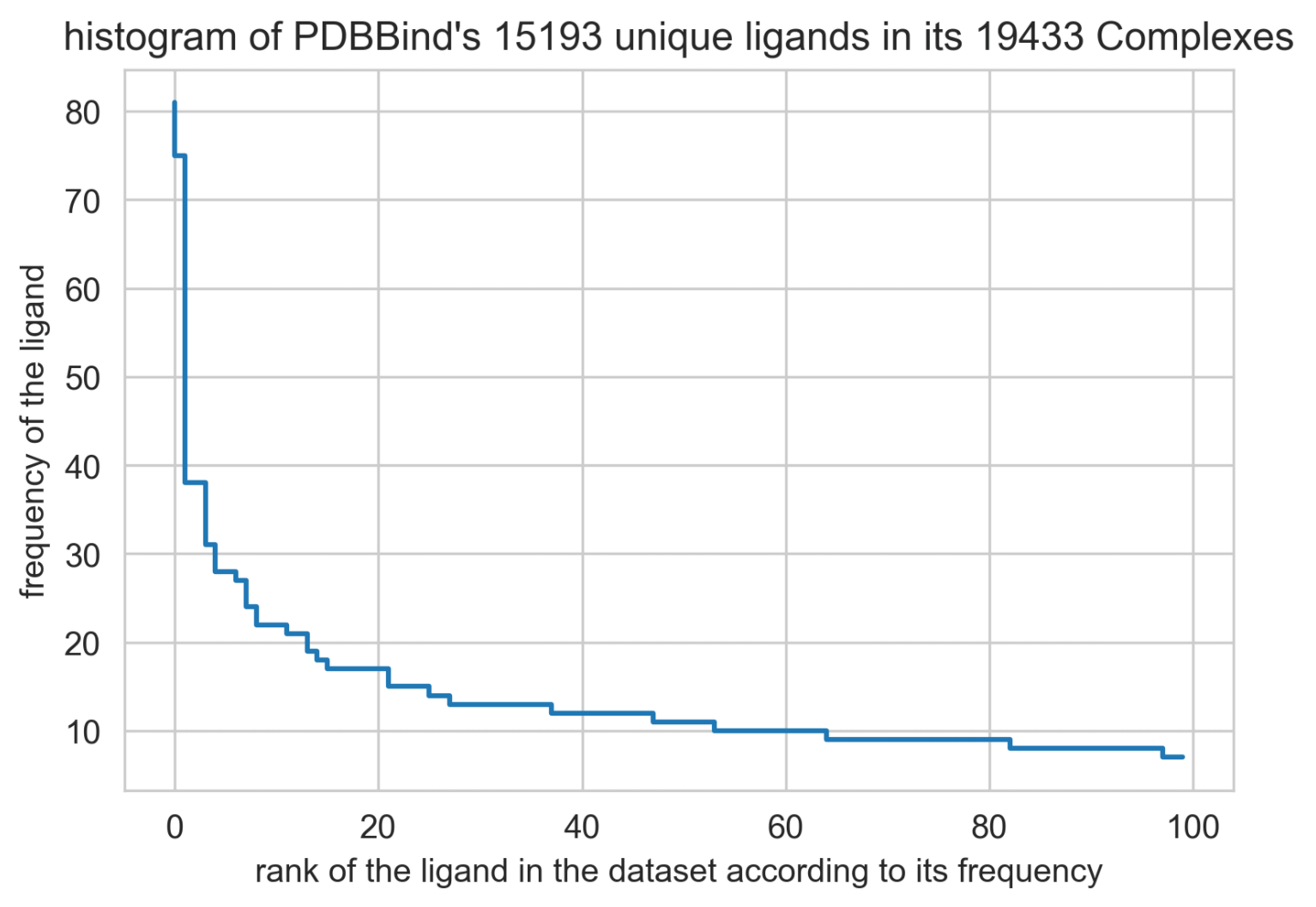}
  \hfill
\includegraphics[width=0.47\textwidth]{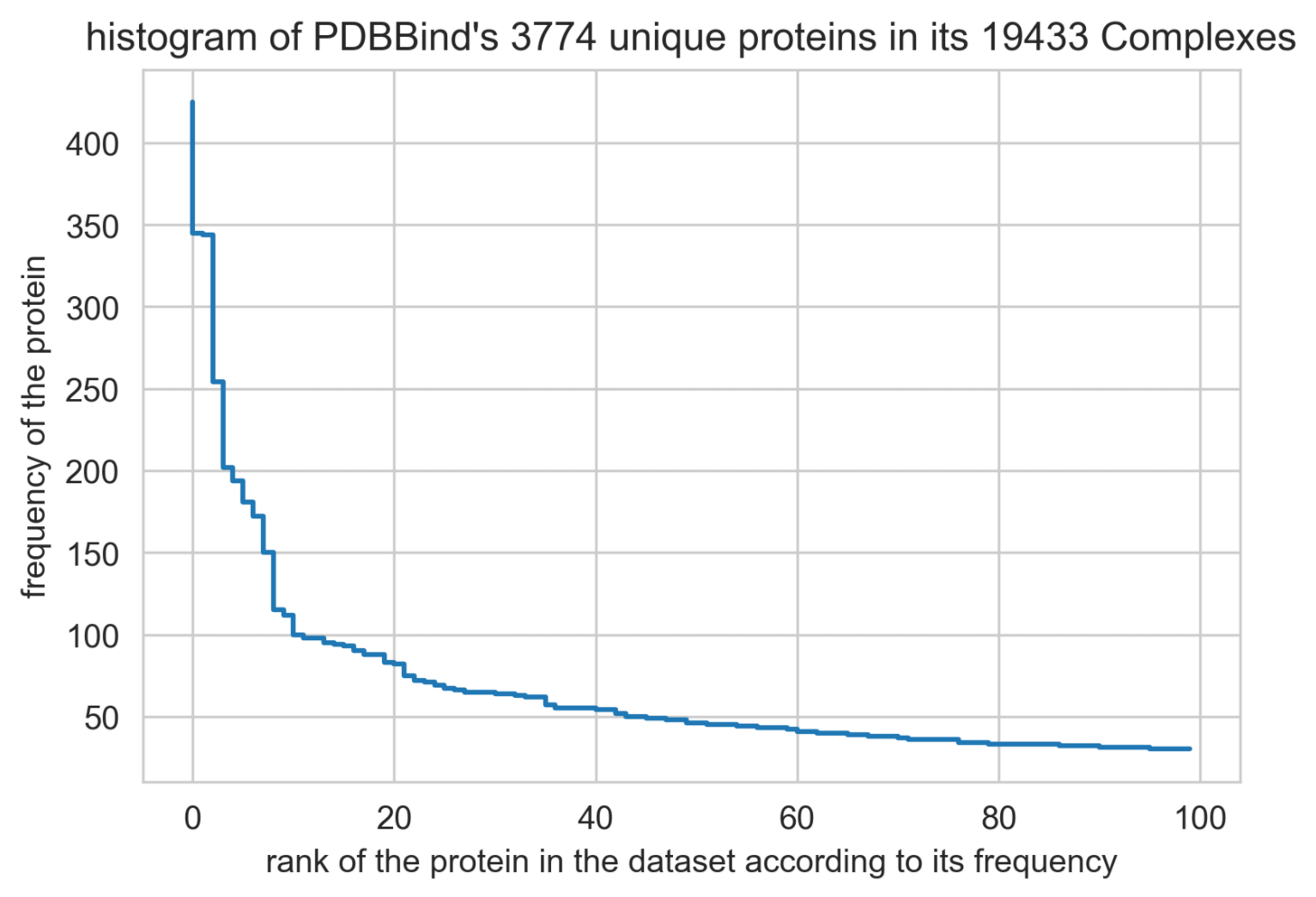}
  \caption{Histograms that show how often each unique ligand and receptor appears in the PDBBind dataset.} \label{dataset-histograms}

\end{figure}

\section{Implementation Details}\label{appendix:implementation}

\textbf{Featurization.}
For the $\alpha$-carbons in the receptor graph, we use the residue type as a feature. The edges have two attributes. Firstly, the interatomic distances encoded with Gaussian basis functions with 15 different variances—secondly, the local frame orientation encodings as they are employed by \citet{Jumper2021} and \cite{ganea2021independent}.

In the ligand, the edges have features that are encoded in the same fashion as for the receptor. Meanwhile, the atoms have the following features: atomic number; chirality; degree; formal charge; implicit valence; the number of connected hydrogens; the number of radical electrons; hybridization type; whether or not it is in an aromatic ring; in how many rings it is; and finally, 6 features for whether or not it is in a ring of size 3, 4, 5, 6, 7, or 8.

\textbf{GLIDE Runtime.}
While baselines like GNINA, SMINA, and QVina-W can leverage multiple CPU cores while making the predictions for a single complex, this is not the case for GLIDE, which only uses a single thread when processing a complex (this and the following information is described here \url{https://www.schrodinger.com/kb/1165}). GLIDE has an application that supports starting multiple processes with each processing a different complex in parallel and for distributing these processes across CPU cores. However, each process also requires a separate software license.

\textbf{Further Hyperparameters.}
We use a learning rate of $10^{-4}$ for \textsc{EquiBind} and $3 \times 10^{-4}$ for \textsc{EquiBind}-R. The learning rate is reduced by a factor of $0.6$ after 60 epochs of no improvement in our main validation criterion, which is the percentage of predicted validation set complexes with an RMSD  better than 2  \AA{}. The models with the best validation score are then tested on the time-based test set.

\begin{table}[htpb]
\caption[Searc]{Search space for all \textsc{EquiBind} models through which we searched to obtain a strong performance on the validation set. The final parameters for the standard \textsc{EquiBind} model are marked in \textbf{bold}.}
\label{search-space}
\begin{center}
\begin{small}
\begin{sc}
\begin{tabular}{lc}
\toprule
Parameter & Search Space  \\    
\midrule
LAS DG step size $\eta$ & 1, 0.01, \textbf{0.001}, 0.0001\\
LAS DG number of steps $T$ & \textbf{1}, 5, 10\\
optimal transport loss weight & 0, 0.1, 0.5, \textbf{1}, 2, 10\\
intersection loss weight & 0, 0.1, 1, \textbf{3}, 10, 50, 100 \\
propagation depth & [ 5, 7, \textbf{8}] \\
intersection $\sigma$ &  8  (based on loss on val-set)\\
intersection $\gamma$ &  8 (based on loss on val-set) \\
kabsch RMSD loss weight & 0, \textbf{1} \\
hidden dimension &  32, \textbf{64}, 100  \\
non linearities & \textbf{leaky-ReLU}, ReLU, SELU \\
learning rates & 0.0009, 0.0003, \textbf{0.0001}, 0.00007\\
dropout & 0, 0.05, \textbf{0.1}, 0.2 \\
num attention heads & 10, 20, \textbf{30}, 50, 100 \\
normalization & \textbf{batchnorm}, layernorm, graphnorm \\
\bottomrule
\end{tabular}
\end{sc}
\end{small}
\end{center}
\vskip -0.1in
\end{table}

\end{document}